\documentstyle[12pt]{article}

\begin{document}

\pagestyle{empty}

\begin{flushright}
{\bf McGILL-00-03}\\
{\bf UA/NPPS-04-00}
\end{flushright}

\vglue 2cm

\begin{center} \begin{Large} \begin{bf} 
Next-to-Leading Order Corrections to Heavy Flavour Production in
Longitudinally Polarized Photon-Nucleon Collisions
\end{bf} \end{Large} \end{center}
\vglue 0.35cm
{\begin{center} 
Z.\ Merebashvili$^{a,*,1}$,
A.P.\ Contogouris$^{a,b,2}$ and 
G.\ Grispos$^{b,3}$ \end{center}}
\parbox{6.4in}{\leftskip=1.0pc
{\it a.\ Department of Physics, McGill University, Montreal,
Qc., H3A 2T8, Canada}\\
\vglue -0.25cm
{\it b.\ Nuclear and Particle Physics, University of Athens, 
Athens 15771, Greece} 
}
\begin{center}
\vglue 1.0cm
\begin{bf} ABSTRACT \end{bf}
\end{center}
\vglue 1.0cm
{\rightskip=1.5pc
\leftskip=1.5pc
\tenrm\baselineskip=12pt
 \noindent
A complete next-to-leading order calculation of longitudinally polarized
heavy quark photoproduction
is presented. All results of the purturbative calculation are given in
detail. For reactions and
energies of interest cross sections differential in the transverse
momentum and rapidity of the
heavy quark, total cross sections and the corresponding asymmetries are
given. Errors in the
asymmetries are estimated and the possibility to distinguish between
various scerarios of the
polarized gluon distribution is discussed. Our results are compared with
other related publications.
%\vglue 2cm
%PACS number(s): 13.60.-r, 13.60.Hb, 13.88.+e
}

\renewcommand{\thefootnote}{*}
\footnotetext{Present address: High Energy Physics Institute,
Tbilisi State University, University St. 9, 380086 Tbilisi, 
Republic of Georgia.}
\renewcommand{\thefootnote}{\arabic{footnote}}
\addtocounter{footnote}{1}
\footnotetext{e-mail: mereb@sun20.hepi.edu.ge}
\addtocounter{footnote}{1}
\footnotetext{e-mail: apcont@physics.mcgill.ca, acontog@cc.uoa.gr}
\addtocounter{footnote}{1}
\footnotetext{e-mail: ggrispos@cc.uoa.gr}

\newpage

\pagestyle{plain}
\setcounter{page}{1}

%\vglue .3cm
\renewcommand{\theequation}{1.\arabic{equation}}

\begin{center}\begin{large}\begin{bf}
I. INTRODUCTION
\end{bf}\end{large}\end{center}
\vglue .3cm

Deep inelastic scattering of longitudinally polarized particles
has provided important information on the spin structure of the
nucleon. However, the size and shape of the polarized gluon
distribution $\Delta g$ in the proton remains an essential
problem. Significant
progress requires experiments on reactions with longitudinally
polarized particles dominated by subprocesses with initial gluons.
Such a reaction is
\begin{equation}
\label{photpr}
\vec{\gamma}+\vec{p}\rightarrow Q \bar{(Q)} + X,
\end{equation}
where $Q \bar{(Q)}$ denotes heavy quark (antiquark); this is
dominated by
\begin{equation}
\label{gamglu}
\vec{\gamma}+\vec{g}\rightarrow Q + \bar{Q}
\end{equation}
An experiment closely related to (\ref{photpr}) is soon going to take
place $[$\ref{r1}$]$
and there are more than one proposals $[$\ref{r1_1}$]$.

At the Born level, (\ref{photpr}) has been studied long ago
$[$\ref{r3},\ref{r3_1}$]$. However, the importance of knowing the
next-to-leading order corrections (NLOC) cannot be overemphasized.
This work presents detailed results on a NLOC calculation.

It should be noted that NLOC for (\ref{photpr}) have already been
published $[$\ref{r3_2}$]$.
We believe, however, that in view of the importance of (1.1), an independent
determination of NLOC in a different regularization approach (see
below) is in order.
Extensive comparisons with the calculation of $[$\ref{r3_2}$]$, as well
as certain
differences in our view regarding certain questions will be also reported.

At NLO, apart from the loop and gluon Bremsstrahlung (Brems)
contributions to (\ref{gamglu}), the subprocesses
\begin{equation}
\label{gamqua}
\vec{\gamma}+\vec{q}(\vec{\bar{q}})\rightarrow Q + \bar{Q}
+ q (\bar{q}),
\end{equation}
where $q$ denotes a light quark, should also be taken into account.

We note that the Abelian part of NLOC for (1.4) provides the
corrections to 
\begin{equation}
\vec{\gamma}+\vec{\gamma}\rightarrow Q + \bar{Q}
\end{equation}
This part has already been determined $[$\ref{r4},\ref{r5}$]$.
NLOC to (1.4) are of interest in themselves in connection with
Higgs boson searches when the Higgs mass is in the range of 90
to 160 GeV.

The loop and $2\rightarrow 3$ parton graphs involved in NLOC
introduce ultraviolet (UV), infrared (IR) and collinear singularities,
which are eliminated by working in $n=4-2\varepsilon$ dimensions.
For polarized reactions this requires extension of the Dirac matrix
$\gamma_5$ in $n\neq 4$ dimensions. Unless otherwise stated, we work
in the scheme of dimensional reduction (RD), which simplifies
the calculation of the traces. Certain subtleties of RD have been
discussed in $[$\ref{r4}$]$ and are mentioned below. Furthermore,
we use parton distributions whose evolution, via 2-loop anomalous
dimensions, is determined in a scheme different from RD. This
necessitates the addition to our perturbative results of certain
conversion terms.

In all the above contributions the photon interacts in a direct way.
In addition, there are also resolved contributions, in which it
interacts through its partonic constituents; in fact, strictly speaking,
at NLO, scheme independent cross sections arise only by adding them. At this
moment a complete calculation of the resolved contributions is not possible,
and we will be limited in giving an estimate. 

The paper is organized as follows. 
Sect.~II contains our general procedures, 
Sect.~III discusses the loop contributions to the photon-gluon fusion
subprocess 
and Sect.~IV the corresponding Brems ones. 
Sect.~V presents analytic results on the
subprocess (1.3). 
In Sect.~VI we derive the nesessary formulas for calculating various
physical observables. 
Sect.~VII presents our numerical results and discusses
the possibility to distinguish between three sets differing essentially in
the polarized gluon distribution function $\Delta g$. 
Sect.~VIII deals with our comparison with $[$\ref{r3_2}$]$, as well as
with $[$\ref{smith}$]$. 
Sect.~IX presents our conclusions. 
Finally, in three Appendices we present results
completing our determination of NLOC.

%\newpage
\renewcommand{\theequation}{2.\arabic{equation}}
\setcounter{equation}{0}
\vglue 1cm
\begin{center}\begin{large}\begin{bf}
II. GENERAL PROCEDURES
\end{bf}\end{large}\end{center}
\vglue .3cm

The Born and the loop contributions to
$\gamma+g\rightarrow Q + \bar{Q}$ are shown in Fig.~1. With
the 4-momenta $p_i, i=1,...,4,$ as indicated and with $m$ the
heavy quark mass we define:
\begin{equation}
s\equiv (p_1+p_2)^2, {\rm \hspace{.3in}}  t\equiv T-m^2
\equiv (p_1-p_3)^2-m^2,
{\rm \hspace{.3in}}  u\equiv U-m^2\equiv (p_2-p_3)^2-m^2
\end{equation}
Let $M_{i}(\lambda_1,\lambda_2)$ the amplitude of any of the
contributing graphs, where $\lambda_1, \lambda_2$ the helicities
of the initial partons; our polarized cross sections correspond
to the quantities:
\begin{equation}
\frac{1}{2} \Sigma [M_{i}(++) \stackrel{\ast}{M}_{j}(++) -
M_{i}(+-) \stackrel{\ast}{M}_{j}(+-)]
\end{equation}
where $\Sigma$ denotes summation over the helicities and colors
of the final particles and average over the colors of the
initial. For the determination of the asymmetries we need also
the unpolarized cross sections, which correspond to the average
of $M_{i}(++) \stackrel{\ast}{M}_{j}(++)$ and
$M_{i}(+-) \stackrel{\ast}{M}_{j}(+-)$.

We also introduce
\begin{equation}
v\equiv 1+t/s {\rm \hspace{.8in}}  w\equiv -u/(s+t)
\end{equation}
To reduce the length of the subsequent expressions we will
make use of the results presented in $[$\ref{r4}$]$. Thus
our leading-order (LO) polarized and unpolarized cross
sections are
\begin{equation}
[\Delta]\frac{d\sigma_{\rm LO}^{\gamma g}}{dvdw} = 
\kappa C_{F} [\Delta]\frac{d\sigma_{\rm LO}}{dvdw}
\end{equation}
where $\kappa \equiv \alpha_s/8\alpha e_{Q}^2$ and
$[\Delta]d\sigma_{\rm LO}/dvdw$ the corresponding [polarized] unpolarized
cross sections
for $\gamma \gamma \rightarrow Q\bar{Q}$ (Eq.~(9) of
$[$\ref{r4}$]$). For later use we note that
$[\Delta]d\sigma_{\rm LO}/dvdw$ are proportional to:
\[
\Delta B(s,t,u)=\frac{1}{s}\left(-\frac{t^2+u^2}{tu} +
2\frac{sm^2}{tu}\left(\frac{s^2}{tu}-2\right)\right)
\]
and (see also $[$\ref{smith}$]$) 
\begin{equation}
\label{e2.5}
B(s,t,u)=\frac{1}{s}\left(\frac{t^2+u^2}{tu} +
4\frac{sm^2}{tu}\left(1-\frac{sm^2}{tu}\right)\right)
\end{equation}

In determining the loop contributions, the renormalization
of the heavy quark mass and wave function were carried on
shell, as in $[$\ref{r4}$]$, i.e. the renormalized heavy
quark self-energy $\Sigma_{r}(p)$ was taken to satisfy
at $p^2=m^2$:
\begin{equation}
\label{e2.6}
\Sigma_{r}(p)=0 {\rm \hspace{.6in}}  \frac{\partial}{\partial
p}\Sigma_{r}(p)=0
\end{equation}
This determines the mass and wave function renormalization
constants $Z_m$ and $Z_2$ $[$\ref{r4}$]$.

Dimensional reduction does not automatically satisfy the Ward identity
\[  Z_1=Z_2,    \]
where $Z_1$ is the renormalization constant for the vertex of
the graph Fig.~1(d). This requires the introduction of proper
finite counterterm, of which the form is given in $[$\ref{r4}$]$.

In the present case charge renormalization is also required.
Defining
\begin{equation}
\label{e2.7}
C_{\varepsilon}(m)\equiv\frac{\Gamma(1+\varepsilon)}{(4\pi)^2}
\left(\frac{4\pi\mu^2}{m^2}\right)^\varepsilon ,
\end{equation}
let $g_{0}(g)$ be the bare (renormalized) coupling, $Z_g=g_0/g$ the
charge renormalization constant and $b=(11N_C-2N_{lf})/6$, where
$N_{lf}$ is the number of light flavors. We take
\begin{equation}
\label{e2.8}
Z_g=1-\frac{g^2}{\varepsilon}
\left\{C_{\varepsilon}(M)b-\frac{1}{3} C_{\varepsilon}(m)\right\}
\end{equation}
where $M$ is a regularization mass. In this scheme the contribution
of a heavy quark loop in the gluon self-energy is subtracted out,
i.e. the heavy quark is decoupled $[$\ref{r5_1},\ref{smith}$]$. This is
consistent with parton
distributions $\Delta F_{a/p}(x,Q^2)$ of which the evolution is
determined from split functions involving only light quarks, as
is the case of $\Delta F_{a/p}$ used below.

Finally, the renormalization of the $gQ\bar{Q}$ vertex was
carried using the Slavnov-Taylor identities $[$\ref{r6}$]$.

\renewcommand{\theequation}{3.\arabic{equation}}
\setcounter{equation}{0}
\vglue 1cm
\begin{center}\begin{large}\begin{bf}
III. LOOP CONTRIBUTIONS
\end{bf}\end{large}\end{center}
\vglue .3cm

The loop graphs contributing to (1.2) are depicted in Fig.~1. The
integrals for the Abelian type of graphs (a)-(e) were calculated in
$[$\ref{r4}$]$. The non-Abelian graphs (g) and (h) introduce
tensor integrals of the form
\[
\int\!\! \frac{d^n q}{(2\pi)^n}
\frac{q^{\mu},q^\mu q^\nu}
{q^2(q-p_2)^2[(q+p_4-p_2)^2-m^2]}     \]
and
\[    \int\!\! \frac{d^n q}{(2\pi)^n}
\frac{q^{\mu},q^\mu q^\nu,q^\mu q^\nu q^\rho}
{q^2(q-p_2)^2[(q+p_4-p_2)^2-m^2][(q-p_3)^2-m^2]}
\]
As in $[$\ref{r4}$]$, using Passarino-Veltman techniques
$[$\ref{r7}$]$, we reduce them to scalar ones; those can be
found in $[$\ref{r8}$]$.

The contributions presented below include the
$t\leftrightarrow u$ crossing symmetric of Fig.~1 plus UV
counterterms; thus they contain no UV singularities.

The graphs (a)-(e) give:
\begin{equation}
\label{e3.1}
\frac{d\sigma_{a-e}^{\gamma g}}{dvdw}=
\kappa C_F\frac{d\sigma_{\rm vse}}{dvdw}-\frac{N_C}{2}
\frac{d\sigma_{a-e}}{dv}\delta(1-w)
\end{equation}
where $d\sigma_{\rm vse}/dvdw$ is given in Eq.(16) of $[$\ref{r4}$]$ and
\begin{eqnarray}
\label{e3.2}
\nonumber
%\lefteqn{  
\frac{d\sigma_{a-e}}{dv}&=&K_L\{ 2\tilde{A}_1 
[2(\zeta(2)-{\rm Li}_2(\frac
{T}{m^2}))(1+3\frac{m^2}{t}) - \ln(\frac{-t}{m^2})(1+\frac{m^2}{T})
+2] + \tilde{A}_2 \ln(\frac{-t}{m^2})                   \\
& & + \tilde{A}_3 ({\rm Li}_2(\frac{T}{m^2})
-\zeta(2)) + \tilde{A}_4 + (t\leftrightarrow u) \}
\end{eqnarray}
with
\[
K_L\equiv \frac{1}{8s}\alpha\alpha_s^2 e_Q^2
\]
Here and subsequently the polarized cross sections are given by
(3.1) and (3.2) with $\Delta d\sigma/dvdw$, $\Delta d\sigma/dv$
and $\Delta \tilde{A}_1$, i=1,...,4, replacing the corresponding
unpolarized quantities. The $[\Delta] \tilde{A}_i$ are given in
Appendix A.

Graph (f) contributes:
\begin{equation}
\label{e3.3}
\frac{d\sigma_f}{dvdw}=\kappa (C_F-\frac{N_C}{2})\frac{d\sigma_{\rm box}}
{dvdw}
\end{equation}
with $d\sigma_{\rm box}/dvdw$ in Eq.(22) of $[$\ref{r4}$]$.

Turning to the non-Abelian graphs, (g) gives
\begin{equation}
\label{e3.4}
\frac{d\sigma_g^{\gamma g}}{dvdw}=-2\pi\alpha_s C_{\varepsilon}(m)
N_C\frac{d\sigma_{\rm LO}^{\gamma g}}{dvdw}\left(\frac{1}
{\varepsilon^2}+\frac{4}{\varepsilon}\right)-
\frac{N_C}{2}\left(\frac{d\tilde{\sigma}_g}{dv}+
\frac{d\sigma_g}{dv}\right)\delta(1-w),
\end{equation}
where
\begin{equation}
\label{e3.5}
\frac{d\tilde{\sigma}_g}{dv}=2F(\varepsilon)\left\{
-A_1\frac{1}{\varepsilon}\ln(\frac{-t}{m^2})+{A'}_1\left[
\frac{1}{\varepsilon^2}-\frac{2}{\varepsilon}-
\frac{2}{\varepsilon}\ln(\frac{-t}{m^2})\right]
+(t\leftrightarrow u)\right\}
\end{equation}
with
\begin{equation}
\label{e3.6}
F(\varepsilon)\equiv K_{L}
\mu^{2\varepsilon} \left(\frac{4\pi\mu}{m}\right)^{2\varepsilon}
\frac{\Gamma(1+\varepsilon)}{\Gamma(1-\varepsilon)}
\left(\frac{s\mu^2}{tu-sm^2}\right)^{\varepsilon}
\end{equation}
and
\begin{eqnarray}
\label{e3.7}
\nonumber
\frac{d\sigma_g}{dv}=K_L \{ 2A_1[
{\rm Li}_2(\frac{T}{m^2})+\ln^2(\frac{-t}{m^2})-2]
+{A'}_1 [4{\rm Li}_2(\frac{T}{m^2})+4\ln^2(\frac{-t}{m^2})]    \\
+ {A'}_2\ln(\frac{-t}{m^2}) + {A'}_3 + (t\leftrightarrow u)\}
\end{eqnarray}
In (3.5) and (3.7), $[\Delta] A_1$ are given in App. B of
$[$\ref{r4}$]$ and $[\Delta] {A'}_i$, i=1,2,3, in App. A of this paper.

The contribution of the graph (h) is:
\begin{equation}
\label{e3.8}
\frac{d\sigma_h^{\gamma g}}{dvdw}=-\frac{N_C}{2}\left\{
4\pi\alpha_s C_{\varepsilon}(m)\frac{3}{\varepsilon^2}
\frac{d\sigma_{\rm LO}^{\gamma g}}{dvdw} + \left(
\frac{d\tilde{\sigma}_h}{dv}+
\frac{d\sigma_h}{dv}\right)\delta(1-w)\right\},
\end{equation}
where
\begin{equation}
\label{e3.9}
\frac{d\tilde{\sigma}_h}{dv}=2F(\varepsilon)\left\{
-A_1\frac{1}{\varepsilon}\left[\ln(\frac{-t}{m^2})+
2\ln(\frac{-u}{m^2})\right]
+{A'}_1\left[-\frac{1}{\varepsilon^2}+\frac{2}{\varepsilon}+
\frac{2}{\varepsilon}\ln(\frac{-t}{m^2})\right]
+(t\leftrightarrow u)\right\}
\end{equation}
and
\begin{eqnarray}
\label{e3.10}
\nonumber
\frac{d\sigma_h}{dv}=K_L \{ A_1[
-\frac{35}{4}\zeta(2)-{\rm Li}_2(\frac{T}{m^2})+4\ln(\frac{-t}{m^2})
\ln(\frac{-u}{m^2})-\ln^2(\frac{-t}{m^2})]
+{B'}_1 {\rm Li}_2(\frac{T}{m^2})                    \\
\nonumber
+ ({B'}_2+{A'}_1) \zeta(2) + {B'}_3 
\ln^2(\frac{-t}{m^2}) + {B'}_4 \ln(\frac{-t}{m^2}) + 
{B'}_5 \ln(\frac{-t}{m^2})\ln(\frac{-u}{m^2})         \\
+ {B'}_6 + (t\leftrightarrow u) \}
\end{eqnarray}
The coefficients $[\Delta] {B'}_i$, i=1,...,6, are given in App. A.

Finally, after cancellation of the UV singularities, graph (i) does not
contribute.

We remark that regarding the terms $1/\varepsilon^2$, the
contributions of the graphs (g) and (h) taken separately are not
proportional to the Born $d\sigma_{\rm LO}^{\gamma g}/dvdw$; only
their sum is proportional to the Born. The same holds regarding
the terms $1/\varepsilon$.

\renewcommand{\theequation}{4.\arabic{equation}}
\setcounter{equation}{0}
\vglue 1cm
\begin{center}\begin{large}\begin{bf}
IV. GLUON BREMS CONTRIBUTIONS
\end{bf}\end{large}\end{center}
\vglue .3cm

In this chapter we present complete analytic results for the NLOC
arising from Brems. To the best of our knowledge, in relation with
heavy quark production, such results have not so far been presented.
 
With $k$ the 4-momentum of the emitted gluon we introduce also
\begin{equation}
\label{e4.1}
s_2\equiv (k+p_4)^2-m^2 = s+t+u = sv(1-w)
\end{equation}
The Brems graphs contributing to the NLOC of (1.2) are shown in
Fig.~2A. The squared sum of the corresponding amplitudes (plus those
obtained via $p_1\leftrightarrow p_2$) after summing over final
spins and colors and averaging over initial colors is given by
\begin{equation}
\label{e4.2}
4m^2|M_{2\leftrightarrow 3}^{\gamma g}|^2=K_B(\varepsilon)\left(
\frac{C_F}{2}G^{\gamma\gamma} - \frac{N_C}{16}G^{\gamma g}\right)
\end{equation}
where $G^{\gamma\gamma}$ is the quantity in the square bracket of
Eq.(24) of $[$\ref{r4}$]$ (plus $p_1\leftrightarrow p_2$),
$G^{\gamma g}$ has the expansion
\begin{eqnarray}
\label{e4.3}
\nonumber \lefteqn{ 
G^{\gamma g} = e_1 
+ \frac{e_2}{p_2\cdot p_4}
+ \frac{e_3}{p_1\cdot p_4^{\,\,2}} 
+ \frac{e_4}{p_1\cdot p_4} 
+ \frac{\tilde{e}_5}{p_3\cdot k} 
+ \frac{\tilde{e}_{6}}{p_1\cdot p_4\,\, p_3\cdot k} 
+ \frac{e_7}{p_1\cdot p_4\,\, p_2\cdot p_4} 
+ e_{8}\frac{p_2\cdot k}{p_1\cdot p_4}          }  \\
\nonumber
&+& \frac{\tilde{e}_{9}}{p_2\cdot p_4\,\, p_3\cdot k} 
+ e_{10}\frac{p_2\cdot k}{p_3\cdot k} 
+ \frac{e_{11}}{p_1\cdot p_4^{\,\,2}\,\, p_3\cdot k}
+ f_1\frac{p_3\cdot k}{p_2\cdot p_4}
+ f_2\frac{p_3\cdot k^{\,\,2}}{p_2\cdot p_4}
+ \tilde{f}_3\frac{p_3\cdot k^{\,\,2}}{p_2\cdot k} + 
\tilde{f}_4\frac{p_3\cdot k}{p_2\cdot k}      \\
\nonumber &+& \frac{\tilde{f}_5}{p_2\cdot k}
+ \frac{\tilde{f}_{6}}{p_1\cdot p_4\,\, p_2\cdot k}
+ \frac{\tilde{f}_7}{p_2\cdot k^{\,\,2}}                      
+ \frac{\tilde{f}_{8}}{p_1\cdot p_4\,\, p_2\cdot k^{\,\,2}}   
+ \frac{\tilde{f}_{9}}{p_2\cdot k\,\, p_3\cdot k} 
+ \frac{\tilde{f}_{10}}{p_1\cdot p_4^{\,\,2}\,\, p_2\cdot k}          \\
&+& \frac{\tilde{f}_{11}}{p_1\cdot p_4^{\,\,2}\,\, p_2\cdot k^{\,\,2}}
\end{eqnarray}
and
\begin{equation}
\label{e4.4}
K_B(\varepsilon)=(4\pi)^3\alpha\alpha_s^2 e_Q^2\mu^{6\varepsilon}
\end{equation}
As in Sect. III, $\Delta |M_{2\rightarrow 3}^{\gamma g}|^2$ is
given by (4.2) and (4.3) with $\Delta G^{\gamma\gamma}$,
$\Delta G^{\gamma g}$, $\Delta e_i$ and $\Delta f_i$, i=1,2,...,13,
replacing the corresponding unpolarized quantities. The coefficients
$[\Delta] e_i$, $[\Delta] f_i$ of (4.3) are given in App. B.

The Brems contribution to $[\Delta]d\sigma/dvdw$ is obtained by
working in the Gottfried-Jackson frame of $\bar{Q}(Q)$ and gluon
(c.m. system of $p_4$ and $k$). Details are given in $[$\ref{r4}$]$.
The terms with coefficients $[\Delta]\tilde{e}_i$, 
$[\Delta]\tilde{f}_i$ in (4.3) give contributions singular at 
$s_2=0\,\,(w=1)$ and must be integrated in $n\neq 4$ dimensions.
In view of the fact that the $2\rightarrow 3$ particle phase space 
is proportional to $s_2^{1-2\varepsilon}$ (Eq. (26) of 
$[$\ref{r4}$]$), the remaining terms can be integrated in 
4 dimensions. The arising integrals are given in $[$\ref{r8}$]$.
Certain terms of special interest not given in $[$\ref{r4}$]$ are
determined in App. C.

Corresponding to the second term in (4.2), with 
$\bar{y}\equiv \sqrt{(t+u)^2-4m^2s}$, $S_2=s_2+m^2$ and 
$x=(1-\beta)/(1+\beta)$, where $\beta=\sqrt{1-4m^2/s}$, the final result is 
\begin{eqnarray}
\label{e4.5}
\nonumber \lefteqn{ \frac{d\sigma_{\rm Br}^{\gamma g}}{dvdw}=
-\frac{K_B(0)}{(4\pi)^3} \frac{N_C}{16} 2\pi 
\frac{vs_2}{8\pi S_2} \{e_1 + 
\frac{2S_2}{s_2(s+u)}\ln\frac{S_2}{m^2}\,\, e_2 
+  \frac{4S_2}{m^2(s+t)^2} e_3 
+ \frac{2S_2}{s_2(s+t)}\ln\frac{S_2}{m^2}\,\, e_4    }                   \\
\nonumber &+&  e_7 I_8 + e_{8} I_{10} + e_{10} I_{16} + 
e_{11} I_{13}(t\leftrightarrow u) + f_1 F_1 + f_2 F_2 \}  \\
\nonumber 
%cons5=
&-& \frac{1}{(1-w)_+} \frac{s_2^2}{S_2} \frac{N_C}{16} F(0)
\{  \frac{2S_2}{s_2\bar{y}} 
\ln\frac{T+U-\bar{y}}{T+U+\bar{y}} \,\, \tilde{e}_5 
+ \tilde{e}_6 I_{11}(t\leftrightarrow u) + \tilde{e}_{9} I_{11} \} \\
\nonumber
%cons1=%should take 2\pi out of itegrals%
&-& \frac{vss_2}{S_2} \frac{N_C}{16} F(0)
\{  \tilde{f}_3 F_{3}^{c} + \tilde{f}_{4} F_{4}^{c} + \tilde{f}_{7} F_{7}^{c} + 
\tilde{f}_{8} F_{8}^{c} + \tilde{f}_{10} F_{10}^{c} + 
\tilde{f}_{11} F_{11}^{c} - (2\ln\frac{s_2}{m^2} + \ln\frac{m^2}{S_2})
(\tilde{f}_3 F_{3}^{s}                                  \\
\nonumber
&+& \tilde{f}_{4} F_{4}^{s} + \tilde{f}_{8} F_{8}^{s} 
+ \tilde{f}_{10} F_{10}^{s} + \tilde{f}_{11} F_{11}^{s})  \}        \\
\nonumber
%cons3=%should take 2\pi out of itegrals%
&-& \frac{1}{(1-w)_+} \frac{s_2^2}{S_2} \frac{N_C}{16} F(0)
\{  \tilde{f}_6 F_{6}^{c} + \tilde{f}_{9} F_{9}^{c}
- (2\ln\frac{sv}{m^2} + \ln\frac{m^2}{S_2})
(\tilde{f}_5 F_{5}^{s} + \tilde{f}_{6} F_{6}^{s} + 
\tilde{f}_{9} F_{9}^{s})    \}                        \\
\nonumber
%cons4=%should take 2\pi out of itegrals%
&+& 2L_+ \frac{s_2^2}{S_2} \frac{N_C}{16} F(0)
\{  \tilde{f}_5 F_{5}^{s} + \tilde{f}_{6} F_{6}^{s} + 
\tilde{f}_{9} F_{9}^{s}    \}                           \\
\nonumber
%CONS2 = 
&+&  8\pi\alpha_s N_C C_\varepsilon(m) [\Delta]\frac{d\sigma_{\rm LO}^
{\gamma g}}{dvdw}
\{ 2\ln^2 \frac{sv}{m^2} - \ln^2(x) +
             \frac{1}{2} \ln^2(\frac{t}{u}x)
             + 2\ln\frac{u}{t} \ln\frac{sv}{m^2}
+ {\rm Li}_2 (1 - \frac{1}{x} \frac{u}{t})              \\
%\nonumber
&-& {\rm Li}_2 (1 - \frac{1}{x} \frac{t}{u}) - 2 \zeta(2)
%CONS5D = 
- \frac{2m^2 - s}{s\beta}
[  (2\ln\frac{sv}{m^2} - \ln(x)) \ln(x) - 
{\rm Li}_2\left(\frac{-4\beta}{(1 - \beta)^2}\right) ] \}
\end{eqnarray}
In Eq. (4.5), the integrals $I_i$ are given 
in the App. C 
of $[$\ref{r4}$]$ and the integrals $F_i$ in the
App. C of this paper. Also, $L_{+}\equiv (\ln(1-w)/(1-w))_{+}$,
which enters through the relation
\begin{equation}
\label{e4.6}
(1-w)^{-1-2\varepsilon}=-\frac{1}{2\varepsilon}\delta(1-w) + 
\frac{1}{(1-w)_{+}} - 2\varepsilon L_{+},
\end{equation}
where the so called "plus" distributions are defined in a usual way:
\begin{equation}
\label{plus}
\int_0^1\,dz\frac{f(w)}{(1-w)_{+}} \equiv
\int_0^1\,dz\frac{f(w)-f(1)}{1-w}.
\end{equation}

For the second term of (4.2), we give the terms 
$\sim 1/\varepsilon^2$ and $1/\varepsilon $, as well: 
\begin{eqnarray}
\label{e4.7}
\nonumber 
\frac{d\sigma_{\rm Br}^{\gamma g, \varepsilon}}{dvdw}&=&
\frac{N_C}{\varepsilon^2}8\pi\alpha_s C_\varepsilon(m)
[\Delta]\frac{d\sigma_{\rm LO}^{\gamma g}}{dvdw} - 
\frac{N_C}{\varepsilon}8\pi\alpha_s C_\varepsilon(m)
[\Delta]\frac{d\sigma_{\rm LO}^{\gamma g}}{dvdw} 
\{ 3 \ln(\frac{-u}{m^2}) - \ln(\frac{-t}{m^2})  \}      \\
\nonumber
&+& \frac{N_C}{\varepsilon}8\pi\alpha_s C_\varepsilon(m)[\Delta]
\frac{d\sigma_{\rm LO}^
{\gamma g}}{dvdw}\frac{2m^2-s}{s\beta} \ln(x)        \\
&-& \frac{1}{\varepsilon}\frac{2sv}{1-vw}F(\varepsilon)[\Delta]P_{gg}^f (x_2)
[\Delta]B(x_2 s, t, x_2 u) 
%%%%%%%%%%%%\left(\frac{S_2m^2}{s_2^2}\right)^{\varepsilon}, 
\end{eqnarray}
where $[\Delta]P_{gg}^f(x)$ is the 4-dimensional $g\rightarrow gg$ split 
function without the $\delta(1-w)$ part, $F(\varepsilon)$, $[\Delta]B(s,t,u)$ 
are given by (3.6) and (2.5) and 
\begin{equation}
\label{e4.8}
x_2=\frac{1-v}{1-vw}
\end{equation}

Addition of loop and Brems contributions cancels the singularities 
$1/\varepsilon^2$ and part of the $1/\varepsilon$. The remaining
$1/\varepsilon$ are cancelled by a factorization counterterm 
corresponding to the final gluon emitted collinearly with the
initial one (Fig.~2A, graph (d)). In the $\overline{MS}$ scheme this
counterterm gives:
\begin{equation}
\label{e4.9}
[\Delta]\frac{d\sigma_{\rm ct}}{dvdw}=\frac{1}{\varepsilon}
\frac{2sv}{1-vw}F(\varepsilon)[\Delta]P_{gg}(x_2)[\Delta]
B(x_2 s, t, x_2 u)\left(\frac{m^2}{M_F^2}\right)^{\varepsilon},
\end{equation}
$[\Delta]P_{gg}(x)$ the $g\rightarrow gg$ split function and  
$M_F$ the factorization scale.

Our cross sections will be convoluted with parton distributions
evolved via two-loop split functions. In $n$ dimensions the split
functions have the form
\begin{equation}
\label{e4.10}
[\Delta]P_{ba}^{n}(x,\varepsilon)=[\Delta]P_{ba}(x)+
\varepsilon [\Delta]P_{ba}^{\varepsilon}(x)
\end{equation}
The polarized split functions have been determined $[$\ref{r9},\ref{r9_1}$]$ 
in the t'Hooft-Veltman scheme $[$\ref{r10}$]$ modified so that
$\Delta P_{qq}^{n}(x,\varepsilon)=P_{qq}^{n}(x,\varepsilon)$.
In this scheme
\begin{equation}
\label{e4.11}
\Delta P_{gg}^{\varepsilon}(x)=4 N_C (1-x) + \frac{1}{6} N_{lf}
\delta(1-x)
\end{equation}
However, our calculations were carried in dimensional reduction
(Sect. I), where
\begin{equation}
\label{e4.12}
\Delta P_{ab}^{\varepsilon}(x)=P_{ab}^{\varepsilon}(x)=0
\end{equation}
Thus, a conversion term $\Delta d\sigma_{\rm conv}/dvdw$ should be
added to our $\Delta d\sigma^{\gamma g}/dvdw$. Conversion terms
are determined from the difference of 
$[\Delta]P_{ab}^{\varepsilon}(x)$ in the two schemes $[$\ref{kam}$]$:
In the present case
\begin{equation}
\label{e4.13}
\Delta\frac{d\sigma_{\rm conv}}{dvdw}=-\frac{2sv}{1-vw}F(0)\Delta
P_{gg}^{\varepsilon}(x_2)\Delta B(x_2 s, t, x_2 u)
\end{equation}
with $\Delta P_{gg}^{\varepsilon}(x)$ given by (4.11).

The unpolarized parton distributions we use were evolved in
the $\overline{MS}$ scheme, where $P_{gg}^{\varepsilon}(x)=0$. 
Thus conversion term is not required.

\renewcommand{\theequation}{5.\arabic{equation}}
\setcounter{equation}{0}
\vglue 1cm
\begin{center}\begin{large}\begin{bf}
V. SUBPROCESS $\gamma q\rightarrow Q\bar{Q}q$
\end{bf}\end{large}\end{center}
\vglue .3cm

The graphs contributing to this subprocess are shown in Fig.~2B. 
The squared sum of the corresponding amplitudes, after summing
over spins and colors and averaging over initial colors, is given by
\begin{equation}
\label{e5.1}
4m^2|M_{2\rightarrow 3}^{\gamma q}|^2=\frac{2}{N_C}(4\pi)^3
\alpha\alpha_s^2(e_Q^2 Q_1 + e_q^2 Q_2 + e_Q e_q Q_3)
\end{equation}
where $e_q$ the charge of the light quark $q$. 
$|M_{2\rightarrow 3}^{\gamma\bar{q}}|^2$ corresponding to 
$\gamma \bar{q}\rightarrow Q\bar{Q}\bar{q}$ is given by the same
expression with an opposite sign of the last term. The quantity
$Q_1$ is given by an expansion similar to (4.3). Next we introduce
\begin{equation}
\label{e5.2}
s_{34}\equiv p_3\cdot p_4 + m^2
\end{equation}
Then $Q_2$ and $Q_3$ are of the form:
\begin{eqnarray}
\label{e5.3}
\nonumber \lefteqn{ 
Q_{2,3} = e_1
+ \frac{e_4}{p_1\cdot p_4} 
+ e_{8}\frac{p_2\cdot k}{p_1\cdot p_4}
+ \tilde{f}_4\frac{p_3\cdot k}{p_2\cdot k}
+ \frac{\tilde{f}_5}{p_2\cdot k}
+ \frac{\tilde{f}_{6}}{p_1\cdot p_4\,\, p_2\cdot k}
+ f_{12}\frac{p_1\cdot k}{s_{34}}
+ f_{13}\frac{p_1\cdot k}{s_{34}^2}       }    \\
\nonumber
&+& \frac{f_{14}}{s_{34}}
+ \frac{\tilde{f}_{15}}{p_1\cdot k}
+ \frac{\tilde{f}_{16}}{s_{34}\,\, p_1\cdot k}
+ \frac{\tilde{f}_{17}}{s_{34}^2\,\, p_1\cdot k}
+ \frac{f_{18}}{s_{34}\,\, p_1\cdot p_4}
+ \frac{\tilde{f}_{19}}{p_1\cdot k\,\, p_2\cdot k}
+ \tilde{f}_{20}\frac{p_3\cdot k}{p_1\cdot k}       \\
&+& \frac{\tilde{f}_{21}}{s_{34}\,\, p_2\cdot k}
\end{eqnarray}
As before, $\Delta |M_{2\rightarrow 3}^{\gamma g}|^2$ is given
by (5.1) and (5.3) with $\Delta Q_r$, r=1,2,3, $\Delta e_i$ and
$\Delta f_j$ replacing $Q_r$, $e_i$ and $f_j$. The coefficients
$[\Delta] e_i$ and $[\Delta] f_j$ are given in the last part
of App. B.

The contribution to $[\Delta] d\sigma/dvdw$ is obtained by
working as in Sect. 4 (c.m. frame of $\bar{Q}(Q)$ and final
light quark). Again the terms with coefficients 
$[\Delta]\tilde{e}_i$ and $[\Delta]\tilde{f}_j$ must be
integrated in $n$ dimensions.

After phase space integrations, we get the following results for the
sets $Q_1,Q_2$ and $Q_3$:
\begin{eqnarray}
\label{e5.4}
\nonumber  \frac{d\sigma_{\rm Br}^{\gamma q, Q_1}}{dvdw}&=&
L e_Q^2 \{e_1+\frac{4S_2}{m^2(s+t)^2} e_3+\frac{2S_2}{s_2(s+t)}
\ln\frac{S_2}{m^2}\,\, e_4 
+e_8 I_{10}+\tilde{f}_4 F_4^c+\tilde{f}_6 F_6^c              \\
&+&\tilde{f}_8 F_8^c+\tilde{f}_{10} F_{10}^c+\tilde{f}_{11} F_{11}^c \}
\end{eqnarray}
\begin{equation}
\label{e5.5}
\nonumber \frac{d\sigma_{\rm Br}^{\gamma q, Q_2}}{dvdw}=
L e_q^2 \{ e_1+f_{12} F_{12}+f_{13} F_{13}+f_{14} F_{14}+
             \tilde{f}_{16} F_{16}^c+\tilde{f}_{17} F_{17}^c+\tilde{f}_{20} 
F_{20}^c \}
\end{equation}
\begin{eqnarray}
\label{e5.6}
\nonumber  \frac{d\sigma_{\rm Br}^{\gamma q, Q_3}}{dvdw}&=&
L e_Q e_q \{ e_1+\frac{2S_2}{s_2(s+t)}\ln\frac{S_2}{m^2}\,\, e_4+e_8 I_{10}+
\tilde{f}_4 F_4^c+\tilde{f}_6 F_6^c+f_{12} F_{12}                \\
&+&f_{14} F_{14}+ \tilde{f}_{16} F_{16}^c+f_{18} F_{18}+\tilde{f}_{19} F_{19}^c+
      \tilde{f}_{20} F_{20}^c+\tilde{f}_{21} F_{21}^c \}
\end{eqnarray}
with
\[   L=\alpha\alpha_s^2\frac{1}{N_C}\frac{v}{8\pi}\frac{\tilde{s}_2}{S_2}
\]

We do not write down expressions containing $1/\varepsilon$ poles coming from 
sets $Q_1$ and $Q_2$ as they are equal with an opposite sign to the 
corresponding counterterms with 
$(S_2m^2/s_2^2)^\varepsilon$ instead of $(m^2/M_F^2)^{\varepsilon}$ (see below).

The singularities arise when the final light quark is collinear
with the initial one ($k\cdot p_2=0$, Fig.~2B, graphs (a), (b))
as well as when the photon is collinear with the light quark
($k\cdot p_1=0$, Fig.~2B, graphs (c), (d)). To eliminate them we
introduce two counterterms. In the second case the counterterm
involves the Born cross section for $\vec{q}\vec{\bar{q}}
\rightarrow Q\bar{Q}$, which is proportional to $[$\ref{cpk}$]$:
\begin{equation}
\label{e5.7}
\Delta B_{q\bar{q}}(s,t,u)=-B_{q\bar{q}}(s,t,u)=-\frac{1}{s}\left(
\frac{t^2+u^2}{s^2} + 2\frac{m^2}{s}\right)
\end{equation}
Moreover, in $n$ dimensions, in the t'Hooft-Veltman scheme:
\[     \Delta P_{gq}^{n}(x,\varepsilon)=C_F\{2-x+2\varepsilon(1-x)\}
{\rm \hspace{.4in}}  
P_{gq}^{n}(x,\varepsilon)=C_F\{\frac{1+(1-x)^2}{x}-\varepsilon x\} \]
and
\[     
\Delta P_{q\gamma}^{n}(x,\varepsilon)=x-\frac{1}{2}-\varepsilon(1-x)
{\rm \hspace{.4in}}  
P_{q\gamma}^{n}(x,\varepsilon)=\frac{x^2+(1-x)^2}{2}-\varepsilon x
(1-x)  \]
Thus, in the $\overline{MS}$ scheme, the first counterterm gives:
\begin{equation}
\label{e5.8}
[\Delta]\frac{d\sigma_{\rm ct}^{(1)}}{dvdw}=\frac{1}{\varepsilon}
\frac{2sv}{1-vw}F(\varepsilon)[\Delta]P_{gq}(x_2)[\Delta]
B(x_2 s, t, x_2 u)\left(\frac{m^2}{M_F^2}\right)^{\varepsilon},
\end{equation}
where $F(\varepsilon)$, $x_2$ and $[\Delta]B$ given by (3.6),
(4.8) and (2.5), and the second gives:
\begin{equation}
\label{e5.9}
[\Delta]\frac{d\sigma_{\rm ct}^{(2)}}{dvdw}=\frac{1}{\varepsilon}
\frac{16s}{9}F(\varepsilon)\frac{e_q^2}{e_Q^2}
[\Delta]P_{q\gamma}(w)[\Delta]
B_{q\bar{q}}(ws, wt, u)\left(\frac{m^2}{M_F^2}\right)^{\varepsilon}
\end{equation}

Although not necessary, it is now customary and even advantageous
$[$\ref{smith}$]$
to average the unpolarized cross section over $n-2$ spin degrees of
freedom for every incoming boson. This convention is employed when fitting
the unpolarized structure functions.
As a result, for the unpolarized case, the
r.h. side of (5.8)
should be multiplied by $(1+\varepsilon)$ and of (5.9) by $(1-\varepsilon)$. 

We note that, upon integration, the singular terms in
$[\Delta]Q_3$ cancel out, as they should since there is no
counterterm proportional to $e_q e_Q$.

Conversion terms are also needed in the present case. Along
the lines of the previous section:
\begin{equation}
\label{e5.10}
[\Delta]\frac{d\sigma_{\rm conv}^{(1)}}{dvdw}=-
\frac{2sv}{1-vw}F(0)[\Delta]P_{gq}^{\varepsilon}(x_2)[\Delta]
B(x_2 s, t, x_2 u)
\end{equation}
and
\begin{equation}
\label{e5.11}
[\Delta]\frac{d\sigma_{\rm conv}^{(2)}}{dvdw}=-\frac{16s}{9}
F(0)\frac{e_q^2}{e_Q^2}
[\Delta]P_{q\gamma}^{\varepsilon}(w)[\Delta]B_{q\bar{q}}(ws, wt, u)
\end{equation}

Finally, we have carried our analytical calculations using REDUCE
$[$\ref{r19}$]$ and
to some extent FORM $[$\ref{r20}$]$.

\renewcommand{\theequation}{6.\arabic{equation}}
\setcounter{equation}{0}
\vglue 1cm
\begin{center}\begin{large}\begin{bf}
VI. PHYSICAL CROSS SECTIONS
\end{bf}\end{large}\end{center}
\vglue .3cm                                                                     

Here we present the necessary formulas needed for calculation
of the differential and total cross sections for the physical process
$\gamma p\rightarrow Q+X$.
This includes derivation of physical cross sections
for both components of the reaction, i.e. pointlike and resolved,
the latter to leading order (LO).
Note we always observe a heavy quark in the final state.
Capital letters in this chapter refer to the kinematic
variables of the physical process and
small letters to those of the subprocess.
Starting with the pointlike component,
the total cross section for the reaction (\ref{photpr}) can be
written
\begin{equation}
\label{tot}
[\Delta] \sigma_{\gamma p}(S)=\int_{x_{min}}^{1}dx
\,\,
[\Delta]f_{b/p}(x,Q^2) [\Delta]\hat{\sigma}_{\gamma b}(s);
\end{equation}
$b$ denotes the corresponding parton, 
$f_{b/p}$ its probability distribution and 
\begin{equation}
           s=x S,  {\rm \hspace{.4in}}  x_{min}=4m^2/S.
\end{equation}
The total partonic cross section $[\Delta]\hat{\sigma}_{\gamma b}(s)$ can
be
calculated straightforwardly by
\begin{equation}
[\Delta]\hat{\sigma}_{\gamma b}(s)=\int_{v_{min}}^{v_{max}} dv 
\int_{w_{min}}^{1} dw\,\,[\Delta]\frac{d\hat{\sigma}_{\gamma
b}(s,v,w)}{dvdw},
\end{equation}
where
\begin{equation}
v_{max/min}=\frac{1}{2} (1\pm \beta), {\rm \hspace{.4in}} 
w_{min}=\frac{m^2}{sv(1-v)}.
\end{equation}

To derive the transverse momentum differential cross
section we note that the transverse momentum $p_T$ of a heavy quark is
invariant under boosts along the beam axis;
also, that our heavy quark rapidities are defined with respect
to the {\it photon}. Consequently we have:
\begin{equation}
\label{difpt}
[\Delta] \frac{d\sigma_{\gamma p}(S,p_T)}{dp_T}=
\int_{x_{min}(p_T)}^{1} dx \,\,
[\Delta]f_{b/p}(x,Q^2) [\Delta]\frac{d\hat{\sigma}_{\gamma
b}(s,p_T)}{dp_T},
\end{equation}
with
\begin{equation}
x_{min}(p_T)=4(p_T^2+m^2)/S,
\end{equation}
\begin{equation}
\label{second}
[\Delta]\frac{d\hat{\sigma}_{\gamma b}(s,p_T)}{dp_T}=
\int_{y_{min}}^{y_{max}}
dy \,\, \frac{2p_T}{sv} [\Delta]\frac{d\hat{\sigma}_{\gamma b}
(s,v,w)}{dvdw}.
\end{equation}
The integration limits on the c.m. rapidity $y$ are 
\[      y_{max}=-y_{min}=\ln( \sqrt{w_m^{-1}}+
                          \sqrt{w_m^{-1}-1} ),  \]
with $w_m\equiv 4(p_T^2+m^2)/s$.
Integration over rapidities in (\ref{second}) is not well defined for
"plus" distributions (given in (\ref{plus})) in the partonic cross
section. The problem is solved with a change of variables; One needs to
consider an integration contour for a heavy quark rapidity
$y$ and split it into two parts that no overlapping (i.e. double counting)
occurs. Formally one would have
\begin{equation}
\label{contour}
\int_{-y_{max}}^{y_{max}}dyf(y)=\int_{-y_0}^{y_{max}}dyf(y\rightarrow
-y) + \int_{y_0}^{y_{max}}dyf(y).
\end{equation}
In
particular, the splitting point $y_0$ must be the point where
function $w=f(y)$ has a minimum. We find 
\[    y_0=\ln( 2\sqrt{(p_T^2+m^2)/s} )   \]
and the general relation between the "new" variable $w$ and the old
variable $y$
is
\begin{equation}
\label{newold}
    {\rm {e}}_{1,2}^{-y}=\frac{1}{ 2\sqrt{\frac{p_T^2+m^2}{s}} }
  \left\{ 1\pm \sqrt{1-\frac{4(p_T^2+m^2)}{sw}} \right\}.   
\end{equation}
The correct sign in (\ref{newold}) is different in different integration
regions, e.g. in the region $[y_0,y_{max}]$ the function ${\rm {e}}^{-y}$
decreases when
one goes from $y_0$ to $y_{max}$, thus minus sign in (\ref{newold}).
Similarly, we find that for the first term of (\ref{contour}) the sign for
${\rm {e}}^{-y}$ in (\ref{newold}) should be positive.
The resulting expression  for the $p_T$ differential
cross section reads:
\begin{equation}
\label{finpt}
[\Delta]\frac{d\hat{\sigma}_{\gamma b}(s,p_T)}{dp_T}=\int_{w_m}^{1}
\frac{dw}{w}\frac{2p_T}{s \sqrt{1-w_m/w}}
\left\{ [\Delta]\frac{d\hat{\sigma}}{dvdw}(v=v_{+})+
[\Delta]\frac{d\hat{\sigma}}{dvdw}(v=v_{-}) \right\},
\end{equation}
where
\begin{equation}
v_{\pm}=\frac{1}{2}(1\pm \sqrt{1-w_m/x}).
\end{equation}

However, even the expression (\ref{finpt}) is not well suited for numerical
integration. One notices that there is a numerically divergent
(though analytically integrable) square root in the
denominator. The singularity comes from the lower limit $x_{min}(p_T)$
of the $x$ integration. To avoid this minor problem one more change of
variables is necessary. Instead of the old variables $x,w$ we introduce
the new variables $z,w'$ through the relations
\begin{equation}
z=\sqrt{1-\frac{x_{min}(p_T)}{wx}}, {\rm \hspace{.4in}}  w'=w.
\end{equation}
To correctly define integration limits for the new variables one has to
perform a nontrivial mapping.
Finally we obtain:
\begin{equation}
\label{final}
[\Delta]\frac{d\sigma_{\gamma b}(S,p_T)}{dp_T}=
\frac{4p_T}{S x_{min}(p_T)} \int_0^{z_{m}} dz
\int_{w'_m}^{1}
dw\, [\Delta] F_{b/p}(x,Q^2)
\left\{ [\Delta]\frac{d\hat{\sigma}}{dvdw}(v_{+})+
[\Delta]\frac{d\hat{\sigma}}{dvdw}(v_{-}) \right\}.
\end{equation}
$F_{b/p}(x,Q^2)$ is a momentum distribution and 
\begin{equation}
z_{m}=\sqrt{1-x_{min}(p_T)},        {\rm \hspace{.2in}}
w'_{m}=\frac{x_{min}(p_T)}{1-z^2},  {\rm \hspace{.2in}}
x=\frac{w'_m}{w}.
\end{equation}

For the rapidity $Y$ fixed one gets the
following expression:
\begin{equation}
\label{dify}
[\Delta] \frac{d\sigma_{\gamma p}(S,Y)}{dY}=
\int_{x_{min}(Y)}^{1} dx \,\,
[\Delta]f_{b/p}(x,Q^2) [\Delta]\frac{d\hat{\sigma}_{\gamma b}(s,y)}{dy},
\end{equation}
with
\begin{equation}
x_{min}(Y)={\rm e}^{-Y}/(\sqrt{S}/m-{\rm e}^{Y}),
\end{equation}
and
\begin{equation}
\label{finy}
[\Delta]\frac{d\hat{\sigma}_{\gamma b}(s,y)}{dy}=\int_{w_{min}}^{1}
\frac{2wdw}{({\rm e}^y+w {\rm e}^{-y})^2}
[\Delta]\frac{d\hat{\sigma}}{dvdw}
\end{equation}
\[
w_{min}=\frac{{\rm e}^y}{\sqrt{s}/m-{\rm e}^{-y}},
{\rm \hspace{.2in}}  v=\frac{1}{1+w{\rm e}^{-2y}},
{\rm \hspace{.2in}}  y=Y+\frac{1}{2}\ln{x}.
\]

Finally we turn to the resolved LO photon contributions. We define the doubly
differential cross section $d\sigma/dYdp_T$ for the
$2\rightarrow 2$ subprocess:
\begin{equation}
[\Delta] \frac{d\sigma_{\gamma p}}{dY dp_T}=2p_T
\int_{x_{1,min}}^1 dx_1  \frac{ [\Delta] F_{a/\gamma}(x_1,Q^2)
[\Delta] F_{b/p}(x_2^0,Q^2)}
{x_1-{\rm {e}}^{Y}A} [\Delta] \hat{\sigma} _{ab}(s,x_1,x_2^0),
\end{equation}
where
\begin{equation}
x_2^0\equiv \frac{x_1 {\rm {e}}^{-Y} A}{x_1-{\rm {e}}^{Y}
A},
{\rm \hspace{.2in}}  A\equiv (\frac{p_T^2+m^2}{S})^{1/2},
{\rm \hspace{.2in}}  s=x_1 x_2^0 S.
\end{equation}

The expressions (6.1) - (6.19) give all the formulas we have used.

\renewcommand{\theequation}{7.\arabic{equation}}
\setcounter{equation}{0}
\vglue 1cm
\begin{center}\begin{large}\begin{bf}
VII. NUMERICAL RESULTS
\end{bf}\end{large}\end{center}
\vglue .3cm

We present results for Q=c-quark ($m_c=1.5$ GeV) at $\sqrt{S_{\gamma
p}}\equiv \sqrt{S}=10$ GeV,
relevant to the experiments $[$\ref{r1}$]$ and $[$\ref{r1_1}$, (a)]$ and
$\sqrt{S}=100$ GeV,
as well as for Q=b-quark ($m_b=5$ GeV) at $\sqrt{S}=100$ GeV; the later
energy is relevant to HERA.
Higher HERA energies are not considered as the cross sections become too
small. The effect of changing
$m_c$ is also considered.

We use the NLO sets of polarized parton distributions of $[$\ref{r16}$]$,
which can be characterized in terms of the polarized gluon distribution
$\Delta g\left( x\right) $ as follows:
\begin{itemize}
\item[ ] Set A: $\Delta g\left( x\right) >0$ and relatively large
\item[ ] Set B: $\Delta g\left( x\right) >0$ and small
\item[ ] Set C: $\Delta g\left( x\right) $ changing sign; $\Delta g\left(
x\right) <0$ for $x>0.1$.
\end{itemize}
Notice that in the presented results also the LO contribution is
convoluted with
NLO distributions; in this way we believe that e.g. the magnitude of
K-factors more properly reflects the NLO subprocess terms. Also, we use
throughout the NLO expression of $\alpha_s \left( \mu \right)$ with
the values for the QCD scale $\Lambda$, flavor thresholds and number of
active flavors 
$N_{lf}=N-1$ that match the definitions corresponding to heavy
quark decoupling.
We note that in $[$\ref{r3_2}$]$ the above values were taken to match the 
definitions for the respective parton distributions.
However, we have explicitly verified that this amounts to a negligible
change in the final numerical results.
Note, in (2.8) we take $M=m$.

At this moment there is no experimental information on the polarized
photon structure functions $\Delta F_{q/ \gamma}$ and  $\Delta F_{g/
\gamma}$, which determine the resolved $\gamma$ contributions. To estimate
them we have used the LO maximal and minimal saturation sets of
$[$\ref{r17}$]$, as well as the sets of $[$\ref{r18}$]$, which belong to
the class of the so-called asymptotic solutions. The two sets of
$[$\ref{r17}$]$ give contributions differing little, with the maximal
saturation one slightly exceeding; the results presented below correspond
to this set. The largest resolved contributions come from $[$\ref{r18}$]$.

In Figs.~3I, II and III, at $\sqrt{S}=10$ and 100 GeV we present
quantities
related with the differential cross sections $\Delta d\sigma /d p_T$,
where $p_T=p_{3T}$ (Fig.~1), versus $x_T\equiv 2p_T/\sqrt{S}$. Measurement
of such cross sections at $\sqrt{S}\approx 10$ GeV may be carried in
(a) of $[$\ref{r1_1}$]$. Here we use the renormalization and factorization
scale $\mu =M_f= \left( p_T^2 +m^2 \right) ^{1/2}$.

In the parts (a) of Figs.~3I, II and III we present the NLO and LO
(denoted by a $*$) contributions to the physical differential cross
section for
sets A, B and C of $[$\ref{r16}$]$. For set B we also present the
contribution of subprocess (1.3) and of the resolved photon.

In the parts (b) of the same Figs we present the asymmetries
\begin{equation}
A_{LL} (p_T)= \frac{\Delta d\sigma /d p_T}{d\sigma /d p_T}
\end{equation}
The unpolarized distributions are the most recent set, CTEQ5
$[$\ref{cteq}$]$. In $A_{LL}$ the resolved $\gamma$ contributions have
been left out since they are small and what is presently known does not
permit a completely scheme independent calculation.
The errors have been estimated using
\begin{equation}
\label{errors}
\delta A_{LL} =  \frac{1}{P_B P_T \sqrt{L \sigma \epsilon } }
\end{equation}
At $\sqrt{S}=10$ GeV we use the conditions of $[$\ref{r1}$]$ ($P_B=80 \%$,
$P_T=25 \%  $, $L=2$ $fb^{-1}$, c-quark detection efficiency $\epsilon _c
=0.014 $) and unpolarized cross section $\sigma$ integrated over a bin of
$x_T$ corresponding to $\Delta p_T=0.5$ GeV. At $\sqrt{S}=100$ GeV we use
$P_B=P_T=70 \%  $,
$L=100$ $pb^{-1}$, $\epsilon _c =0.15 $, for b-quark $\epsilon _b =0.05 $
and $\sigma$ integrated over a bin corresponding to $\Delta p_T=5$ GeV.

Figs.~3I(a) and 3II(a) show that between $\sqrt{S}=10$ and $100$ GeV the
shape of the LO $\Delta d\sigma_{LO} /d p_T$ and NLO $\Delta d\sigma /d
p_T$ varies dramatically;
this also holds for the
K-factor, $K=\Delta d\sigma /d p_T / \Delta d\sigma_{LO} /d p_T$.

Most impotrant is the possibility to distinguish between sets A, B and
C. Fig.~3I(b) shows that at $\sqrt{S}=10$ GeV near $x_T=0.3$ one can
distinguish A and C and perhaps all A, B, C.
Figs.~3II(b) and 3III(b) show that at $\sqrt{S} =100$ GeV the best range
is $0.2 \leq x_T \leq 0.3$;
and for $Q=c$ one may distinguish all A, B, C, but for $Q=b$ only A and C.

In Figs.~4I, II and III we present rapidity distributions. Here we use
$\mu =M_f =2m$. The presented differential cross sections are analogous to
those of Figs.~3I, II and III and 
\begin{equation}
A_{LL} (Y)= \frac{\Delta d\sigma /dY}{d\sigma /dY}
\end{equation}
The errors have been estimated using (\ref{errors}) where now the
unpolarized cross sections $\sigma$
are integrated over a bin $\Delta Y=1$.

Fig.~4I(b) shows that at $\sqrt{S}=10$ GeV the region $1.25 \leq Y \leq
1.5$ is the best to distinguish
set C from A or B. Fig.~4II(b) shows that at $\sqrt{S}=100$ GeV for
c-quark, $A_{LL} (Y)$ has become
too small. At $Y \approx -1$ it seems one can distinguish all A, B, C,
but $\Delta d\sigma /dY$ is
small for all sets (Fig.~4II(a)). Perhaps more promising is the range $0
\leq Y \leq 1$, where one
can distinguish C from A or B. Finally Fig.~4III(b) shows that detection
of b-quark is not useful due
to large errors ($\epsilon _b$ small).

Figs.~5I,II and III present integrated cross sections $\Delta \sigma$ and
the corresponding asymmetries $A_{LL}=\Delta \sigma/\sigma$ versus the
c.m. energy $\sqrt{S}$. The scale is again $\mu=M=2m$.

Comparison of Figs.~5I and 5II shows that at the two different ranges of
$\sqrt{S}$ the changes in the shapes and signs of $\Delta \sigma$ and
$A_{LL}$ is again dramatic; clearly the same holds for the corresponding
$K$-factors, $K=\Delta \sigma_{NLO}/\Delta \sigma_{LO}$.

In Fig.~5I(b) the error (at $\sqrt{S}$=10 GeV) is estimated using again in
(\ref{errors}) the conditions of $[$\ref{r1}$]$. Under these conditions we
conclude that sets A and C can be distinguished, but not sets A and B or B
and C. The proposed SLAC experiment $[$\ref{r1_1}a$]$, which amounts to
better conditions, and will give results at somewhat lower $\sqrt{S}$, may
distinguish also B and C.

In Figs.~5II(b) and 5III(b) the errors (at $\sqrt{S}$=100 GeV) have been
estimated using again the values of $P_B,P_T,L,\varepsilon_c$ and
$\varepsilon_b$ stated after Eq.~(\ref{errors}). For $c$-quark, $\mid
A_{LL} \mid$ are very small due to relatively large unpolarized cross
sections $\sigma$. For the same reason, however, the error $\delta A_{LL}$
is not very large, so set C can be distinguished from A or B. For
$b$-quark, due to a combination of small $\varepsilon_b$ and rather small
$\sigma$, the error is very large and precludes any useful information on
$\Delta g$.

Finally, Fig.~6, for the integrated NLO cross sections $\Delta \sigma$ and
$\sigma$ and for the asymmetries $A_{LL}=\Delta \sigma/\sigma$, shows the
effect of changing the $c$-quark mass $m_c$ (part (a)) and the scales
$\mu,\, M_f$ (part (b)). The results refer to set B of
$[$\ref{r16}$]$. E.g. regarding $\Delta \sigma$, in Fig.~6(a) we define
\begin{equation}
R_m=\frac{\Delta \sigma(m_c)-\Delta \sigma(1.5\,\,{\rm {GeV}})}{\Delta
\sigma(1.5\,\,{\rm {GeV}})},
\end{equation}
and in Fig.~6(b), keeping $\mu=M_f$, we define
\begin{equation}
R_{SC}=\frac{\Delta \sigma(\mu)-\Delta \sigma(2 m_c)}{\Delta \sigma(2 m_c)
};
\end{equation}
similarly for $\sigma$ and $A_{LL}$. Fig.~6(a) shows that at the lower
$\sqrt{S}$ the effect of changing $m_c$ is more pronounced.

\renewcommand{\theequation}{8.\arabic{equation}}
\setcounter{equation}{0}
\vglue 1cm
\begin{center}\begin{large}\begin{bf}
VIII. COMPARISON WITH OTHER PUBLICATIONS
\end{bf}\end{large}\end{center}
\vglue .3cm

Figs.~3II(b) and 3III(b) show that at small $x_T$, $A_{LL} (p_T)$ is
small; the same holds for 
$A_{LL} (Y)$ of Fig.~4II(b). This may lead one to conclude that HERA is
rather useless in
specifying $\Delta g$ $[$\ref{r3_2}$]$. However, it may not be so. On the
basis of Figs.~3II(b) and III(b), reconstruct events and select only
those with, say, $x_T > 0.2$,
i.e. carry integrations of $\Delta d\sigma /d p_T $ over some cut phase
space. This
may well enhance the resulting $A_{LL} $ $[$\ref{our}$]$.. Of course, an estimate of the
corresponding
errors is required to reach a definite conclusion. 

Finally, since we present analytic results for the unpolarized cross
section as well,
we will compare with similar results of  $[$\ref{smith}$]$ ("soft" part,
Eq.~(2.24) of $[$\ref{smith}$]$);
here the relevant part is the last three lines of
Eq.~(4.5). Ref.~$[$\ref{smith}$]$ uses the phase
space slicing method, which separates the soft and hard gluon parts via a
cut parameter $\Delta$.
The formal relation with our approach is
\begin{equation}
\frac{sv} {m^2}\rightarrow \Delta ;
\end{equation}
the necessary framework to relate these two methods is
developed in $[$\ref{bz}$]$.
Now, concerning terms involving $t$ and $u$, we easily see that they are
exactly the same,
except that $t$ and $u$ are interchanged (c.f. our definition,
Eq.~(2.1) with that of $[$\ref{smith}$]$,
Eq.~(2.13)). The only difference seems to arise from the coefficient of
$\zeta (2)$, which is
$-2$ in our case versus $-3/2$ in $[$\ref{smith}$]$. Note, however, that
our coefficient $F(\varepsilon)$
in (3.6) contains $\Gamma (1+\varepsilon )$, which upon expansion in
powers of $\varepsilon $,
gives a term $\left( \varepsilon ^2 /2 \right) \zeta (2)$; this accounts
for the difference. Since
$F(\varepsilon)$ appears both in our loop contribution (3.6) and in our
Brems contribution (4.7),
the overall result is unaffected. To verify our calculation we have
evaluated numerically the NLO $\overline{MS}$ scaling
functions for the partonic
$\gamma g$ cross section, taking into account an additional
"mass" factorization term given in eq.~(6.31) of $[$\ref{r8}$]$,
and
compared it to the corresponding curves of Fig.~5 of $[$\ref{smith}$]$.
We found exact agreement.
We have also explicitly verified that the sum of our non-Abelian loop
contributions and the Brems ones, that are proportional to the Born
contribution, equal analytically the corresponding
"virtual+soft" expression presented in $[$\ref{r3_2}$]$.

\renewcommand{\theequation}{9.\arabic{equation}}
\setcounter{equation}{0}
\vglue 1cm
\begin{center}\begin{large}\begin{bf}
IX. CONCLUSIONS
\end{bf}\end{large}\end{center}
\vglue .3cm

In this paper we have presented the complete analytic results for the
heavy flavor photoproduction for both, longitudinally polarized and
unpolarized initial particles, in a closed form. These include the NLO
contributions of the hard Brems due to the relevant partonic
subprocesses (\ref{gamglu}) and (\ref{gamqua}) that are presented for the
first time in analytic form. We have computed numerically various
total and differential cross sections for the energy ranges of CERN, SLAC
and HERA. We have discussed the possibilities to differentiate between
various scenarios for the polarized gluon distribution $\Delta
g$ and have once more emphasized the way
to enhance the asymmetries for HERA energies by measuring the differential
cross sections with the help of certain acceptance cuts (see also our earlier Ref.
$(a)$ of $[$\ref{our}$]$ on this subject).

\vglue 1cm
\begin{center}\begin{large}\begin{bf}
ACKNOWLEDGEMENTS
\end{bf}\end{large}\end{center}
\vglue .3cm
We thank I.~Bojak for his kind collaboration in doing comparisons.
Thanks are also due to G.~Bunce, D.~de~Florian, B.~Kamal and J.~K{\"o}rner
for discussions, to W.~Vogelsang for discussions and for providing us the
sets of $[$\ref{r20}$]$, to P.~Bosted for several communications, to
A.~Despande for useful information and remarks and to V.~Spanos and 
G.~Veropoulos for participating in part of the calculations. Z.M. would
like to thank the Particle Theory group of the Institut f{\"u}r Physik,
Universit{\"a}t Mainz, for hospitality, where the calculations of the
final parts of this paper were carried out.

\vglue 1cm
\begin{center}\begin{large}\begin{bf}
APPENDIX A
\end{bf}\end{large}\end{center}
\vglue .3cm

\setcounter{equation}{0}
\renewcommand{\theequation}{A\arabic{equation}}

Here we present the coefficients of the loop contributions. In the
following $[\Delta]A_i$, i=1,3, are given in App. B of 
$[$\ref{r4}$]$. For $[\Delta]d\sigma_{a-e}/dv$ given in Eq. (3.2):
\begin{eqnarray}
\nonumber \Delta \tilde{A}_1&=& \Delta A_1;   {\rm \hspace{.2in}}
\Delta \tilde{A}_2=-4[2(7s^2/t^2+8s/t+6)m^2/u+11s^2/tu+24s/u+26t/u+\\
\nonumber & &  {\rm \hspace{1.8in}} 12t^2/su+st/uT-2t^2/sT]m^2/T   \\
\nonumber \Delta \tilde{A}_3&=&\Delta A_3/2;   {\rm \hspace{.2in}}
\Delta \tilde{A}_4=4[(2u/t-s/u)m^2/t-2s/t+2u/s]m^2/T   \\
\nonumber \tilde{A}_1&=& A_1;  {\rm \hspace{.2in}}
\tilde{A}_2=4[(24s/t-2s^2/tT+2s/T+12t/T)m^4/tu+(s/t
       +6t/s+t/T+11)                                   \\
\nonumber  & &  {\rm \hspace{1.0in}} m^2s/uT-2t^2/T^2]         \\
\tilde{A}_3&=& A_3/2;   {\rm \hspace{.2in}}
\tilde{A}_4=-4[4m^4s/ut^2-(2s^2/t^2-s/t+t/T-2)m^2/u-t/T]
\end{eqnarray}

For $[\Delta]d\tilde{\sigma}_g/dv$ and $[\Delta]d\sigma_g/dv$ 
given in Eqs. (3.5) and (3.7):
\begin{eqnarray}
\nonumber \Delta {A'}_1&=& m^2s/t^2                    \\
\nonumber \Delta {A'}_2&=&4[2m^4s^2/Tut^2-(8su/t^2+9u/T-8-t/T
       -8t^2/sT-st/T^2+2t^2u/sT^2)m^2/u]               \\
\nonumber \Delta {A'}_3&=& 4[m^4s/Tut-2(s/t-1-t^2/sT)m^2/t]      \\
\nonumber {A'}_1&=& -m^2s^2/t^2u-1                      \\
\nonumber {A'}_2&=&-4[(16sT/t^2+2s^2/t^2+8-u/T+t/T)m^4/uT+2(s/u
            -2s/T)m^2/t]                                \\
{A'}_3&=& 4[(4sT/t^2+s/t+2)m^4/uT+2m^2s^2/ut^2+3] 
\end{eqnarray}

For $[\Delta]d\tilde{\sigma}_h/dv$ and $[\Delta]d\sigma_h/dv$
given in Eqs. (3.9) and (3.10):
\begin{eqnarray}
\nonumber \Delta {B'}_1&=& 2[4(5/u+1/t)m^4/t+(5/u+7/t+4u/t^2)m^2+
                           (4/u+3/t)(t^2+u^2)/s]            \\
\nonumber \Delta {B'}_2&=&-(15s^2/tu-62)m^2s/tu-4(s^2/tu+8)m^4/tu
                           +13/4(s^2/tu-2)               \\
\nonumber \Delta {B'}_3&=& 2[(3/u+4/t)(u-t)m^2/t+(1/s-1/t)
                          (t^2+u^2)/u]      \\
\nonumber \Delta {B'}_4&=& 4[2(2/u+3/t+u/t^2)m^2-(t^2/u-t+9u+3u^2/t
                           -6tu/T-t^4/T^2u+t^3/T^2)/s]           \\
\nonumber \Delta {B'}_5&=& 8(s^2/tu-3)m^2s/tu               \\
\nonumber \Delta {B'}_6&=& 2[2(1/T+1/U)m^2/s+2(t^2/u^2+u^2/t^2+
               2t/u+2u/t)/s-t/uU-u/tT-                       \\
\nonumber  & &              (t^3/u-2tu+u^3/t)/sTU]m^2    \\
\nonumber {B'}_1&=& 2[4(1/t-3/u)m^4/t-(1-u/t)(1/u+2/t)m^2+
                                      4t/u+2+3u/t]         \\
\nonumber {B'}_2&=& [4(29/u^2-6/tu+29/t^2)m^4-2(1/u^2-12/tu+1/t^2)
                    m^2s-13t/u+4-13u/t]/4                    \\
\nonumber {B'}_3&=& 2[4(1/t-3/u)m^4/t+(2u/t^2-1/t-5/u)m^2-2s/u+u/t]\\ 
{B'}_4&=& -4[2(4/u+1/t)m^2s/t+4t/u-3s/t-2t^2/Tu-t/T-
                                      2t^3/T^2u]         \\
\nonumber {B'}_5&=& -16(1/t^2+1/u^2)m^4                    \\
\nonumber {B'}_6&=& -4[(t/Uu^2+u/Tt^2)m^4-(5s/tu-1/U-1/T-t^2/TUu
                    -u^2/TUt)m^2+2tu/TU] 
\end{eqnarray}

\vglue 1cm
\begin{center}\begin{large}\begin{bf}
APPENDIX B
\end{bf}\end{large}\end{center}
\vglue .3cm

\setcounter{equation}{0}
\renewcommand{\theequation}{B\arabic{equation}}

In this Appendix we list the coefficients of the Brems contributions. 
For $\Delta G^{\gamma g}$ given in Eq.~(4.3) and $
\Delta d\sigma_{\rm Br}^{\gamma g}/dvdw$ 
given in Eq.~(4.5), the coefficients $\Delta e_i$ and $\Delta f_i$ are
\begin{eqnarray}
\nonumber \Delta {e}_1&=& 16[4(s/s_2u - 1/u + 1/t)m^2/t + 
            3s/s_2u - 2/u - s/s_2t - s_2/tu - u/s_2t]            \\
\nonumber \Delta {e}_2&=& 8[8(-s/u - 1 + s/t)m^4/s_2t + 
            2(2s/u - 2 + 4s/t - 3ss_2/t(s+u))m^2/s_2 -        \\
\nonumber & &            4s^2/s_2u + 4s/u - 2s/s_2 -
            u/s_2 + u/t + 2ss_2/tu + 3u/(s+u)]               \\
\nonumber \Delta {e}_3&=& 0;        {\rm \hspace{0.6in}}
\Delta {e}_4 = 8[4(s/u + s/s_2 - t/s_2)m^2/t + 
            ss_2/tu + su/s_2t - 2u/s_2 - 2]               \\
\nonumber \Delta\tilde{e}_5&=& 8[8(t/u + 1)m^4/s_2t - 
         2(s^2/s_2u + s/u + 4t/s_2 - 2s_2/u)m^2/t + st/s_2u +  \\
\nonumber & &  4s/s_2 - 2 + su/s_2t + 2u/t]                   \\
\nonumber \Delta\tilde{e}_6&=& 4[8(s/u - s_2/t + u/t)m^4/s_2 + 
         2(2st/s_2u + s/s_2 - s_2/u + 2 + 2ss_2/tu -        \\
\nonumber & & 2s^2/s_2t)m^2 + s(s^2+u^2)/s_2t]                  \\
\nonumber \Delta e_7&=& 4[8(s_2/u - 1)m^4/t + 
        2(2s/u + 3 - s_2/s - 2s_2^2/tu + 2s/t)m^2s/(s+u) -      \\
\nonumber & & ss_2(s^2 + s_2^2)/t(s+u)u]; {\rm \hspace{0.6in}}
\Delta e_{8} = 16(s_2/u - u/s_2)/t          \\
\nonumber \Delta\tilde{e}_{9}&=& 4[8(t/u + s/t)m^4/s_2 - 
          2(2s^2/s_2u - 2s/u + 2s_2/u - 2 + (st/s_2 + 2su/s_2 -  \\
\nonumber & &  s_2)/(s+u))m^2 - st
          (2s/s_2u - 2/u + (s_2/u + u/s_2)/(s+u))]      \\
\nonumber \Delta e_{10}&=& 16(t/u-u/t)/s_2;   {\rm \hspace{1.2in}}
\Delta e_{11} = - 16m^4s/u                              \\
\nonumber \Delta f_1&=& 16[4(s/u - 1)m^2/s_2t + (3/s_2 + 1/t)s/u];
{\rm \hspace{0.6in}}   \Delta f_2 = 32(s + u)/ts_2u              \\
\nonumber \Delta\tilde{f}_3&=& \Delta f_2;  {\rm \hspace{0.6in}} 
\Delta\tilde{f}_4 = - 32[(2/s_2 + 2/t - s/s_2u)m^2/t + 1/t + 1/s_2] \\
\nonumber \Delta\tilde{f}_5&=& 8[2(s/s_2u + s/tu - 2u/t^2 + 
         s/t(s+u) - 4ss_2/t^2(s+u))m^2 - s^2/s_2u +           \\
\nonumber & &  2s/u - 2s/s_2 + 
         2 - 2u/s_2 + ss_2/tu + s_2/t + (u + 2s_2u/t)/(s+u)]  \\
\nonumber \Delta\tilde{f}_6&=& 4[2(2s^2/s_2u + s/u + s_2/u + 
         3s/s_2 + (s + 2su/s_2 - 
         3s_2 - 4s(s_2^2+                                   \\
\nonumber & &     u^2)/tu)/(s+u))m^2 + s^2/s_2 + 2su/s_2 + 
                  2u^2/s_2 - 2s^2/t - 4su/t +               \\
\nonumber & &     2s_2^2/t - 2u^2/t + (s_2^2 + u^2)/(s+u)]     \\
\nonumber \Delta\tilde{f}_7&=& 0;  {\rm \hspace{0.6in}}
\nonumber \Delta\tilde{f}_8 = 0                         \\
\nonumber \Delta\tilde{f}_9&=& 4[2(2s/u - 2s_2^2/tu + s_2/t - 
               (ss_2/u + 2su/t - s_2u/t)/(s+u))m^2 + s_2 -       \\
\nonumber & &  2t - s_2^2/t - u^2/t + (s_2^2+u^2)/(s+u)] \\
\Delta\tilde{f}_{10}&=& - 8(2s + t)m^2; 
{\rm \hspace{0.6in}}     \Delta\tilde{f}_{11} = 0
\end{eqnarray}
For $G^{\gamma g}$ given in Eq.~(4.3) and $d\sigma_{\rm Br}^{\gamma g}/dvdw$ 
given in Eq.~(4.5), the coefficients $e_i$ and $f_i$ are
\begin{eqnarray}
\nonumber {e}_1&=& 16[4(1/s_2u+1/s_2t-1/t^2)m^2 - 3s/s_2u+2/u-1/s_2-s_2/tu-1/t] 
       \\
\nonumber {e}_2&=& 8[-16(1/t+1/u)m^6/s_2t - 8(3t/s_2u+2/u+2/s_2-
       1/(s+u))m^4/t +         \\
\nonumber & & 2(s^2/s_2u-8s/u+4s/s_2-1+2u/s_2-
       3s_2/t+u/t)m^2/(s+u) + u/s_2-              \\
\nonumber & & 4st/s_2u-2s/s_2+(2s_2-2u+2ss_2^2/tu+s_2u/t)/(s+u)]         \\
\nonumber {e}_3&=& -16m^2           \\
\nonumber {e}_4&=& -8[8m^4(1/u+1/s_2) - 2m^2(s/u-s/s_2-2s_2/u-2u/s_2)+ (s_2+u)
                                                           (2+     \\
\nonumber & &       2u/s_2+2s/s_2-s/u+s/s_2)]/t               \\
\nonumber \tilde{e}_5&=& 8[8(s/s_2-1)m^4/tu - 2m^2(s/s_2u+2/u-5/s_2+3/t-u/ts_2)
        +s^2/s_2u-s/u-  \\
\nonumber & &  s/s_2+4+su/ts_2-2s_2/t+2s^2/ts_2]                 \\
\nonumber \tilde{e}_6&=& 4[16m^6(1/su-1/s_2u-1/ts_2) + 8
        (su-(s-s_2)^2-(s_2-u)^2)m^4/ts_2u + 2m^2(t/u+        \\
\nonumber & &   s/s_2-1+s/t)-s(s^2+u^2)/s_2t]                \\
\nonumber e_7&=& 4[16(s_2/u-1)m^6/st + 8(s^2+ss_2+s_2^2)m^4/tu(s+u1) +
       2m^2(3s/t+(ss_2/u-s_2-                             \\
\nonumber & & 2ss_2/t)/(s+u))-ss_2(s^2+s_2^2)/tu(s+u1)];  {\rm \hspace{0.6in}}
e_{8} = 16(s_2 + u)^2/ts_2u          \\
\nonumber \tilde{e}_{9}&=& 4[16(s/s_2-2+s_2/s-u/s)m^6/tu + 8(s_2/u-
         3st/s_2u-2+u/s_2)m^4/(s+u) + 2m^2           \times     \\
\nonumber & & ((s-s_2)(1/u+1/s_2)-1+2(s_2-s)/(s+u)) + 
(s_2/u-2+u/s_2-2st/s_2u)st/(s+u)]  \\
\nonumber e_{10}&=& -16(1/t-1/u)(s/s_2 - 1);   {\rm \hspace{1.2in}}
e_{11} = 32m^6/u                              \\
\nonumber f_1&=& -16[8m^4/ts_2u+2(2/u+s/tu)m^2/s_2+s/u(1/t+3/s_2)-
     2ss_2/t(s+u)u]             \\
\nonumber f_2&=& 32(1/t+1/s_2)/u;    {\rm \hspace{1.2in}}
\tilde{f}_3= f_2                          \\
\nonumber \tilde{f}_4&=&-32[4m^4/ts_2u+2(1/s_2+1/t)m^2/u-1/s_2+s_2/t(s+u)] \\
\nonumber \tilde{f}_5&=& -8[8m^4(1/s_2u+1/tu+(1/t-2s_2/t^2)/(s+u)) - 4m^2(
         1/t+2s_2/t^2-(st/s_2+                   \\
\nonumber & &      ss_2/t)/(s+u)u) - t/u+2t/s_2
         -s_2^2/tu+(s^2t/s_2u+2u-2s_2u/t)/(s+u)]           \\
\nonumber \tilde{f}_6&=& 4[8m^4(s^2/ts_2u-1/s_2-4/t-(s/t-s_2s/ut-s_2u)/(s+u))
        +4(s^2/s_2+su/s_2-                                  \\
\nonumber & &  2s_2+s-u-s_2(s+s_2)/(s+u))m^2/t
        -(1-(s_2^2+u^2)/(s+u1)t)(s+s_2+u)^2/s_2]                  \\
\nonumber \tilde{f}_7&=& -32(s + u)^2m^2/t^2;  {\rm \hspace{0.6in}}
\nonumber \tilde{f}_8 = 32(s + u)m^2u/t                         \\
\nonumber \tilde{f}_9&=& 4[8m^4(s^2/tu+s_2/u+s_2/t) + 4m^2(
         ss_2/t+s_2-2s) + (s_2/t-2)(s_2^2+u^2-st-      \\
& &  ut)]/(s+u);  {\rm \hspace{0.6in}}
\tilde{f}_{10}= 16m^2(2m^2 + u); 
{\rm \hspace{0.6in}}     \tilde{f}_{11} = -8m^2u^2
\end{eqnarray}

Now we shall write down the coefficients $[\Delta]e_i$ and $[\Delta]f_j$ 
for the subprocess $\gamma q\rightarrow Q\bar{Q}q$. \\
For $Q_1$ we have:
\begin{eqnarray}
\nonumber \Delta {e}_1&=& - 8(2m^2 - t)/t^2, {\rm \hspace{0.2in}}
\Delta {e}_3 = 0, {\rm \hspace{0.2in}}    \Delta {e}_4 = - 8*s/t, 
{\rm \hspace{0.2in}}    \Delta {e}_{8} = 0                    \\
\nonumber
\Delta\tilde{f}_4&=& 8(2m^2 + t)/t^2, {\rm \hspace{0.2in}} 
\Delta\tilde{f}_5 = 4(2(s_2 + s)m^2 - ut)/t^2         \\
\nonumber
\Delta\tilde{f}_6&=& - 2(2m^2 + 2s_2 - t), {\rm \hspace{0.1in}} 
\Delta\tilde{f}_7 = 0,
{\rm \hspace{0.1in}} \Delta\tilde{f}_8 = 0, {\rm \hspace{0.1in}} 
\Delta\tilde{f}_{10} = 2(2s + t)m^2, {\rm \hspace{0.1in}} 
\Delta\tilde{f}_{11} = 0     \\
%unpolarized:
\nonumber
{e}_1&=& 8(2m^2 + t)/t^2, {\rm \hspace{0.2in}} {e}_3=4m^2, {\rm \hspace{0.2in}}
e_4=8(2u + s)/t, {\rm \hspace{0.2in}} e_{8}= - 16/t              \\
\nonumber
\tilde{f}_4&=&8/t, {\rm \hspace{0.2in}} \tilde{f}_5 = - 4(4m^4+4m^2s_2+ut)/t^2 
                                                                       \\
\nonumber
\tilde{f}_6&=&2(4(s_2 + u)m^2 + 2(2s + 3t)s_2 + 8m^4 - 4s_2^2 - 2s^2 - 4st - 
3t^2)/t \\
\nonumber
\tilde{f}_7&=&8(s_2-t)^2m^2/t^2, {\rm \hspace{0.2in}} \tilde{f}_8=-8u(s_2-t)m^2
                                                                 /t   \\
\tilde{f}_{10}&=&-4(m^2+u)m^2, {\rm \hspace{0.2in}} \tilde{f}_{11}=2u^2m^2 
\end{eqnarray}
For $Q_2$:
\begin{eqnarray}
%polarized:
\nonumber \Delta {e}_1&=& - 16/s, {\rm \hspace{0.2in}} \Delta {f}_{12}= - 8/s,
{\rm \hspace{0.2in}} \Delta {f}_{13}= - 8m^2/s, {\rm \hspace{0.2in}} 
\Delta {f}_{14}= - 8(2m^2 + u)/s             \\
\nonumber
\Delta\tilde{f}_{15}&=&8t/s, {\rm \hspace{0.2in}} 
\Delta\tilde{f}_{16}=2(2u(s + t) - 4m^2s + s^2 + 2ts_2)/s          \\
\nonumber
\Delta\tilde{f}_{17}&=&2m^2s, {\rm \hspace{0.2in}} \Delta\tilde{f}_{20}=0   \\
%unpolarized:
\nonumber e_1&=&16/s, {\rm \hspace{0.2in}} {f}_{12}=8/s, {\rm \hspace{0.2in}}
{f}_{13}=8m^2/s, {\rm \hspace{0.2in}} {f}_{14}=8(2m^2 + u)/s       \\
\nonumber
\tilde{f}_{15}&=&8(2m^2 + u)/s, {\rm \hspace{0.2in}} \tilde{f}_{16}=
- 2(4m^2s - 4s_2^2 + 6s_2s + 4s_2t - 3s^2 - 4st - 2t^2)/s                \\
\tilde{f}_{17}&=& 2m^2s, {\rm \hspace{0.2in}} \tilde{f}_{20}=- 16/s
\end{eqnarray}
And, finally, for $Q_3$:
\begin{eqnarray}
%polarized:
\nonumber \Delta {e}_1&=& 8(s_2 - 2s + t)/st, {\rm \hspace{0.2in}} \Delta {e}_4
                                                  =-8, 
{\rm \hspace{0.2in}} \Delta {e}_8=0                       \\
\nonumber
\Delta\tilde{f}_{4}&=& - 8(s_2 - t)/st, {\rm \hspace{0.2in}} 
\Delta\tilde{f}_{5}=4(2(s_2 - t)m^2 + s_2^2 - s_2t - ut)/st           \\
\nonumber
\Delta\tilde{f}_{6}&=& - 2(2(2s + t)m^2 - 2s_2t + t^2)u/(s + t)s,
{\rm \hspace{0.2in}} \Delta f_{12}=8(s_2 - s)/st                 \\
\nonumber
\Delta f_{14}&=&4(2(s_2 - 3s)m^2 - 2su + 2s_2u + ts_2)/st, {\rm \hspace{0.2in}}
\Delta\tilde{f}_{15}=8                         \\
\nonumber
\Delta\tilde{f}_{16}&=& - 2(2(s + 2t)s_2 - 2m^2s - s^2 - 4st - 2t^2)u/(s + t)t 
                                                                        \\
\nonumber
\Delta f_{18}&=& - 2(2((2s + t)s_2 + 2s^2 + 2st)m^2 - (2u + t)s_2t)/(s + t)s  
                                                                        \\
\nonumber
\Delta\tilde{f}_{19}&=& - 2(2(s + t)s_2 - 2m^2s - s^2 + 2ut)s_2/(s + t)t,
{\rm \hspace{0.2in}} \Delta\tilde{f}_{20}=0, {\rm \hspace{0.2in}} 
\Delta\tilde{f}_{21}=0  \\
%unpolarized:$
\nonumber
e_1&=&-8(s_2 - 2s - t)/st, {\rm \hspace{0.2in}} e_4=8(2m^2 + s_2 + u + tu/(s +
                                                                   t))/s,
{\rm \hspace{0.2in}} e_8= - 16/s          \\
\nonumber
\tilde{f}_{4}&=& - 8(s_2 - t)/st, {\rm \hspace{0.2in}} 
\tilde{f}_{5}=4(2(s_2 - t)m^2 + s_2^2 - s_2t - ut)/st         \\
\nonumber
\tilde{f}_{6}&=& - 2(2(4s/t + t/(s+t))m^2 + 2s_2 + 2u + (2s_2t - 4s_2^2 - t^2)/
(s+t))u/s  \\
\nonumber
f_{12}&=& - 8(s_2 - s)/st, {\rm \hspace{0.2in}} 
f_{14}= - 4(2(s_2 - 3s)m^2 + 2s_2u + s_2t - 2su)/st              \\
\nonumber
\tilde{f}_{15}&=& 8(2m^2 + u + ss_2/(s+t))/t                      \\
\nonumber
\tilde{f}_{16}&=& 2(2m^2s - 4s_2^2 + 6s_2s + 4s_2t - 3s^2 - 4st - 2t^2)u/(s + 
                                                               t)t    \\
\nonumber
f_{18}&=& - 2(2(4su/t + 2s - s_2t/(s+t))m^2 + s_2(s_2 + u - 
((s-2s_2)^2+2s_2s)/(s + t)))/s \\
\nonumber
\tilde{f}_{19}&=& 2(2(s + 2t)s_2 - 2m^2s - 4s_2^2 - s^2 - 2st - 2t^2)s_2/(s + 
                                                                  t)t  \\
\tilde{f}_{20}&=& - 16/t, {\rm \hspace{0.2in}} \tilde{f}_{21}= - 16m^2s/t
\end{eqnarray}

\vglue 1cm
\begin{center}\begin{large}\begin{bf}
APPENDIX C
\end{bf}\end{large}\end{center}
\vglue .3cm

\setcounter{equation}{0}
\renewcommand{\theequation}{C\arabic{equation}}

We give here the Brems integrals, $F_i$, $i=1,...,11$, 
appearing in Eq.~(4.5) and also in Eqs.~(5.4) - (5.6).
Define,
\begin{equation}
P_1=u s_2-s (t+2m^2), {\rm \hspace{0.6in}} 
P_2=s ( (t+m^2) (u s_2-m^2 s)-m^2 (s+u)^2 )
\end{equation}

We may now write down the integrals:
\begin{eqnarray}
\nonumber
F_1&=&-\frac{1}{(s+u)^2}[P_1+\frac{s S_2}{s_2} (t+2m^2) \ln\frac{S_2}{m^2}]   \\
\nonumber
F_2&=&\frac{1}{(s+u)^3} [P_1 (\frac{s_2 (s+u) (t+u+2m^2)}{4 S_2}+s (t+2m^2)) + 
         P_2\frac{s_2+2m^2}{2 S_2} + \frac{S_2}{s_2}           \\
\nonumber &\times& (\frac{s^2 (t+2m^2)^2}{2}-P_2
         \frac{m^2}{S_2})\ln\frac{S_2}{m^2} ]                          \\
\nonumber
F_3^s&=&-\frac{s_2 S_2 u^2}{2(s+u)^3}                                    \\
\nonumber
F_3^c&=&\frac{s_2}{2 S_2(s+u)} [ \frac{(t+u)^2}{4}-m^2 s-\frac{t+u+2m^2}{
                                s+u}P_1-\frac{3}{4}\frac{P_1^2}{(s+u)^2} ] 
                                                               \\
\nonumber
F_4^s&=&\frac{S_2 u}{(s+u)^2},{\rm \hspace{0.3in}} F_4^c=\frac{P_1}{(s+u)^2},
{\rm \hspace{0.3in}} F_5^s=-\frac{2S_2}{s_2(s+u)}, {\rm \hspace{0.3in}} 
                                                     F_5^c=0   \\
\nonumber
F_6^s&=&-\frac{4S_2}{s_2ut}, {\rm \hspace{0.6in}}
F_6^c=\frac{4S_2}{s_2ut} [ \ln\frac{S_2}{m^2} + 2\ln\frac{ut}{ss_2+ut} ]     \\
\nonumber
F_7^c&=&-\frac{8S_2^2}{s_2^2(s+u)^2} (1-\varepsilon)                 \\
\nonumber
F_8^s&=&-\frac{16S_2^2}{s_2^2ut(s+u)} 
             \frac{s_2}{2S_2}\frac{s+t}{u^2t^2}(s+u)(ut-2m^2s)         \\
\nonumber
F_8^c&=&\frac{16S_2^2}{s_2^2ut(s+u)} [
             \frac{s_2}{2S_2}\frac{s+t}{u^2t^2}(s+u)(ut-2m^2s)
             (\ln\frac{S_2}{m^2}+2\ln\frac{ut}{ss_2+ut}) -
             \frac{s_2^2}{S_2^2}\frac{2P_2}{u^2t^2} - 1 + \varepsilon ]    \\
\nonumber
F_9^s&=&\frac{4S_2}{s_2^2 u}, {\rm \hspace{0.6in}}
F_9^c=-\frac{4S_2}{s_2^2 u} [\ln\frac{S_2}{m^2} + \ln\frac{u^2}{(s+u)^2}]   \\
\nonumber
F_{10}^s&=&-\frac{8S_2(s+u)}{s_2 u^2 t^2}, {\rm \hspace{0.6in}}
F_{10}^c=\frac{8S_2(s+u)}{s_2 u^2 t^2} [ \ln\frac{S_2}{m^2} +
         2\ln\frac{ut}{ss_2+ut} + \frac{s_2}{m^2}\frac{ut-2m^2s}{ss_2+ut} ]  \\
\nonumber
F_{11}^s&=&-\frac{32S_2^2}{s_2^2 u^2 t^2}
              \frac{s_2^2}{S_2^2 u^2 t^2}[P_2+\frac{S_2}{s_2}(ss_2+ut)(ut-
                                                             2m^2s)]    \\
\nonumber
F_{11}^c&=&\frac{32S_2^2}{s_2^2 u^2 t^2}
             [ \frac{s_2^2}{S_2^2 u^2 t^2}(P_2+\frac{S_2}{s_2}(ss_2+ut)
             (ut-2m^2s))(\ln\frac{S_2}{m^2}+2\ln\frac{ut}{ss_2+ut})
       - \frac{s_2^2}{S_2^2}\frac{8P_2}{u^2t^2}                  \\
&+& \frac{1}{2}\frac{s_2^2}{m^2S_2} - 1 + \varepsilon ] 
\end{eqnarray}
Note that parts of $F_7^c$, $F_8^c$ and $F_{11}^c$ proportional to
$1-\varepsilon$ cancel out exactly in Eq.~(4.5). The singular and finite
parts of these integrals can be found in Appendix C of $[$\ref{r8}$]$; below we 
give the derivation of ${\cal O}(\varepsilon)$ terms.

We use the momentum 
parametrizations of Appendix A of $[$\ref{r4}$]$ and, as in $[$\ref{r8}$]$, we 
denote
\begin{equation}
F_n^{(k,l)}\equiv \int\!\! d\Omega_n (a+b\cos\theta_1)^{-k} (A+B\cos\theta_1+C
\sin\theta_1\cos\theta_2)^{-l}
\end{equation}
where
\begin{equation}
\int\!\! d\Omega_n \equiv \int_0^\pi \!\! d\theta_1
\sin^{1-2\varepsilon}\theta_1 \int_0^\pi\!\!  d\theta_2
\sin^{-2\varepsilon}\theta_2.
\end{equation}
All the above integrals are proportional to 
$1/a^2=1/\omega_k^2\omega_2^2\sim 1/s_2^2\sim 1/(1-w)^2$; since the 
$2\rightarrow 3$ particle phase space is proportional to 
$(1-w)^{1-2\varepsilon}$, in view of Eq.~(4.6), the terms of 
${\cal O}(\varepsilon)$ give 
finite contributions proportional to $\delta(1-w)$.

\vglue 0.5 cm
\noindent Integral $F_7\equiv\int\!\! d\Omega_n/(p_2\cdot k)^2$:

This is of the type $\hat{I}_n^{(2,0)}$ of $[$\ref{r8}$]$. The result is
\begin{equation}
F_7=-\frac{\pi}{a^2}\frac{1}{1+\varepsilon}=-\frac{\pi}{a^2}(1-\varepsilon+
{\cal O}(\varepsilon^2))
\end{equation}

\vglue 0.5 cm
\noindent Integral $F_8\equiv\int\!\! d\Omega_n/(p_2\cdot k)^2(p_1\cdot p_4)$:

This is of the type $\hat{I}_n^{(2,1)}$ of $[$\ref{r8}$]$, and determination 
of the ${\cal O}(\varepsilon)$ term proceeds as follows. First, defining
\begin{equation}
H\equiv A+B\cos\theta_1+C\sin\theta_1\cos\theta_2,
\end{equation}
one can show the identity
\begin{equation}
\frac{1}{(1-\cos\theta_1)H}=\frac{1}{A+B}\left\{\frac{1}{1-\cos\theta_1}+
\frac{B}{H}+\frac{C\sin\theta_1\cos\theta_2}{(1-\cos\theta_1)H}\right\}
\end{equation}
and by repeated application of it: 
\begin{eqnarray}
\nonumber       \lefteqn{
\frac{\sin\theta_1}{(1-\cos\theta_1)^2H}=}          \\
\nonumber  & &  \frac{1}{A+B}\{  \frac{\sin\theta_1}
{(1-\cos\theta_1)^2} + \frac{B}{A+B}[\frac{\sin\theta_1}{(1-\cos\theta_1)} + 
\frac{B\sin\theta_1}{H} - \frac{C(1+\cos\theta_1)\cos\theta_2}{H} ]         \\
\nonumber &-& 
\frac{C\cos\theta_2}{A+B}[\frac{(1+\cos\theta_1)}{(1-\cos\theta_1)}+
\frac{B(1+\cos\theta_1)}{H}] + \frac{C^2}{(A+B)^2}[\frac{\sin^3\theta_1
\cos^2\theta_2}{(1-\cos\theta_1)^2}                                     \\
&+& \frac{B\sin\theta_1(1+\cos\theta_1)\cos^2\theta_2}
{H}-\frac{C(1+\cos\theta_1)^2\cos^3\theta_2}{H}]  \}
\end{eqnarray}

Note that the fifth term vanishes due to the integration over
$\theta_2$. Also, terms with 
only $H$ in the denominators are finite, consequently have no poles and
cannot
produce finite contributions from their ${\cal O}(\varepsilon)$ terms. 
Thus, we are left with the terms 1,2 and 7. After integrating them in
n-dimensions, 
summing and keeping the relevant order $\varepsilon$ terms,
we arrive to the following result:
\begin{equation}
F_8^{\varepsilon}=\frac{\pi}{a^2(A+B)}\varepsilon
\end{equation}

\vglue 0.5 cm
\noindent Integral $F_{11}\equiv\int\!\! d\Omega_n/(p_2\cdot
k)^2(p_1\cdot p_4)^2$:

Proceeding as before, the term of ${\cal O}(\varepsilon)$ is provided by
\begin{equation}
F_{11}^{\varepsilon}=\frac{\pi}{a^2(A+B)^2}\varepsilon
\end{equation}

Finally, we give the Brems integrals, $F_i$, $i=12,...,21$, 
appearing in Eqs.~(5.4) and (5.6). With
\begin{equation}
P_3=2m^2s+u(s-s_2), {\rm \hspace{0.6in}}
Z=4m^2s(s+t)-s_2^2t
\end{equation}
we have:
\begin{eqnarray}
\nonumber
F_{12}&=&-\frac{1}{\bar{y}^2}[P_1(t\leftrightarrow u)+\frac{s}{2}P_3 F_{14}]  \\
\nonumber
F_{13}&=&-\frac{1}{\bar{y}^2}[\frac{2S_2}{sm^2}P_3+P_1(t\leftrightarrow u) 
F_{14}] \\
\nonumber
F_{14}&=&\frac{2S_2}{s_2\bar{y}}\ln\left( \frac{s_2(s-s_2)+2m^2s+s_2\bar{y}}{
                        s_2(s-s_2)+2m^2s-s_2\bar{y}} \right)                  \\
\nonumber
F_{15}^s&=&-\frac{2S_2}{s_2(s+t)}, {\rm \hspace{0.3in}} F_{15}^c=0   \\
\nonumber
F_{16}^s&=&\frac{4S_2}{s_2su}, {\rm \hspace{0.2in}}
F_{16}^c=-\frac{4S_2}{s_2su}[ \ln\frac{S_2}{m^2} + \ln\frac{u^2}{(s+t)^2} ]  \\
\nonumber
F_{17}^s&=&-\frac{8S_2(s+t)}{s_2s^2u^2}, {\rm \hspace{0.2in}}
F_{17}^c=\frac{8S_2(s+t)}{s_2s^2u^2}[ \ln\frac{S_2}{m^2} + 
                       \ln\frac{u^2}{(s+t)^2} - \frac{s_2P_3}{m^2s(s+t)} ]   \\
\nonumber
F_{18}&=&\frac{4S_2}{s_2\sqrt{-tz}}\ln\left( \frac{Z-2m^2s(s+t)+
               s_2\sqrt{-tz}}{Z-2m^2s(s+t)-s_2\sqrt{-tz}} \right)       \\
\nonumber
F_{19}^s&=&-\frac{8S_2}{s_2^2s}, {\rm \hspace{0.2in}}
F_{19}^c=\frac{8S_2}{s_2^2s} \ln\frac{sS_2}{ss_2+ut}          \\ 
\nonumber
F_{20}^s&=&\frac{S_2t}{(s+t)^2}, {\rm \hspace{0.2in}}
F_{20}^c=\frac{P_1(t\leftrightarrow u)}{(s+t)^2}                  \\
F_{21}^s&=&\frac{4S_2}{s_2st}, {\rm \hspace{0.2in}}
F_{21}^c=-\frac{4S_2}{s_2st} [ \ln\frac{S_2}{m^2} + \ln\frac{t^2}{(s+u)^2} ]
\end{eqnarray}

\newpage
%\vglue 1cm
\begin{center}\begin{large}\begin{bf}
REFERENCES
\end{bf}\end{large}\end{center}
\vglue .3cm

   \begin{list}{$[$\arabic{enumi}$]$} 
    {\usecounter{enumi} \setlength{\parsep}{0pt} 
     \setlength{\itemsep}{3pt} \settowidth{\labelwidth}{(99)} 
     \sloppy} 
\item \label{r1} 
G.~Baum et al, COMPASS Collaboration: CERN/SPLC 96-14 and 96-30.
\item \label{r1_1}
(a) R. Arnold, P. Bosted et al, SLAC-PROPOSAL-E156, 1997;
(b) W.-D.~Nowak, DESY 96-095;
(c) A. de Roeck and T. Gehrmann, DESY-Proceedings-1998-1.
\item \label{r3} 
M. Gluck and E. Reya, Z. Phys. {\bf C39}, 569 (1988);
B. Lampe and E. Reya, Phys. reports (in press). The latter is also a
comprehensive
review of polarized particle processes.
\item \label{r3_1}
M. Stratmann and W. Vogelsang, Z. Phys. {\bf C74}, 641 (1997);
A. Watson, {\it ibid} {\bf C12}, 123 (1982).
\item \label{r3_2}
I. Bojak and M. Stratmanm:
(a) Phys. Lett. {\bf B433}, 411 (1998);
(b) Nucl. Phys. {\bf B540}, 345 (1999);
Erratum: {\it ibid} {\bf B569}, 694 (2000).
\item \label{r4} 
B.~Kamal, Z.~Merebashvili and A.P.~Contogouris, Phys. Rev. {\bf D51}, 
4808 (1995); Erratum: {\it ibid} {\bf D55}, 3229 (1997).
\item \label{r5} 
G.~Jikia and A.~Tkabladze, {\it ibid} {\bf D54}, 2030 (1996).
\item \label{smith}
J.~Smith and W.L.~van Neerven, Nucl. Phys. {\bf B374}, 36 (1992).
\item \label{r5_1}
P. Nason et al, {\it ibid} {\bf B327}, 49 (1989).
\item \label{r6} 
T.~Muta, "Foundations of Quantum Chromodynamics" 
(Word Scientific, 1987)
\item \label{r7} 
G.~Passarino and M.~Veltman, Nucl. Phys. {\bf B160}, 151 (1979)
\item \label{r8} 
W.~Beenakker et al, Phys. Rev. {\bf D40}, 54 (1989).
\item \label{r9} 
R.~Mertig and W.~van Neerven, Z. Phys. {\bf C70}, 637 (1996).
\item \label{r9_1} 
W.~Vogelsang, Phys. Rev. {\bf D54}, 2023 (1996);
Nucl. Phys. {\bf B475}, 47 (1996).
\item \label{r10} 
G.'t~Hooft and M.~Veltman, Nucl. Phys. {\bf B44}, 189 (1972).
\item \label{kam} 
B.~Kamal, Phys. Rev. {\bf D53}, 1142 (1996); {\it see also}
A.P.~Contogouris and Z.~Merebashvili, {\it ibid} {\bf D55}, 
2718 (1997).
\item \label{cpk} 
A.P.~Contogouris, S.~Papadopoulos and B.~Kamal, Phys. Lett. 
{\bf B246}, 523, (1990).
\item \label{r19}
A. Hearn, REDUCE User' s Manual Version 3.6
(Rand Corporation, Santa Monica, CA, 1995).
\item \label{r20}
J. Vermaseren, FORM User' s Manual, NIKHEF-H, Amsterdam, 1990.
\item \label{r16} 
T. Gehrmann and W. Stirling, Phys. Rev. {\bf D53}, 6100 (1996).
 \item \label{r17}
M. Gluck and W. Vogelsang, Z. Phys. {\bf C55}, 353 (1992)
and {\bf C57}, 309 (1993);
M. Gluck, M. Stratmann and W. Vogelsang, Phys. Lett. {\bf B187}, 373 (1994).
\item \label{r18} 
J. Hassan and D. Pilling,
Nucl. Phys. {\bf B187}, 563 (1981).
\item \label{cteq}
H. Lai et al, Eur. Phys. J. {\bf C12} (2000) 375.
\item \label{our}
(a) Talk given by Z.~Merebashvili at the International Workshop Spin 99,
Prague,
Czech Republic, 5-11 Sept. 1999, published in Czech. J. Phys, Vol. {\bf 50},
No. S1, 153 (2000); and hep-ph/9911506;
(b) A.P.~Contogouris, Z.~Merebashvili and G.~Grispos, Phys. Lett. {\bf
B482}, 93 (2000).
\item \label{bz}
B.~Kamal and Z.~Merebashvili, Phys. Rev. {\bf D58}, 074005 (1998).
\end{list}

%\newpage
\vglue 3cm
\begin{center}\begin{large}\begin{bf}
FIGURE CAPTIONS
\end{bf}\end{large}\end{center}
\vglue .3cm

\begin{list}{Fig.~\arabic{enumi}.} 
    {\usecounter{enumi} \setlength{\parsep}{0pt} 
     \setlength{\itemsep}{3pt} \settowidth{\labelwidth}{Fig.~9.} 
     \sloppy}  
\item{}
LO and loop graphs. In the loop graphs $p_1 \leftrightarrow p_2$ crossed
ones are not
shown. Note that graph $(i)$, representing gluon, quark and ghost loop,
does not
contribute here.
\item{} 
A) Gluon Brems graphs; $p_1 \leftrightarrow p_2$ crossed ones are not
shown. B) Graphs of the subprocess $\gamma q \rightarrow Q \overline Q q$.
\item{} 
Quantities related with the $p_T$ distributions versus $x_T = 2p_T
/\sqrt{S} $: Parts
(a): Polarized differential cross sections; the LO (Born) ones are
indicated by $*$.
Parts (b): Asymmetries for sets A, B and C. 3I: $Q=c$, $\sqrt{S} =10$
GeV.
3II: $Q=c$, $\sqrt{S} =100$ GeV. 3III: $Q=b$, $\sqrt{S} =100$ GeV. 
\item{} 
Quantities related with the rapidity $Y$ distributions: Parts (a) and
(b), as well as 4I, 4II
and 4III as in Fig. 3.
\item{} 
Quantities related with the integrated cross sections for
$\vec \gamma \vec p \rightarrow Q+X$:
(a) Factors $K= \Delta \sigma / \Delta \sigma _{LO} $ (b) Asymmetries.
\item{} 
At c.m. energies $\sqrt{S}=10$ and 100 GeV, for integrated cross
sections and with solid lines for $\Delta \sigma$, dashed for $\sigma$
and dotted for the asymmetry $A_{LL}$: a) The ratio $R_m$ (see
end of Sect. VI) with $m=m_c$; b) The fractional variation $R_{sc}$
with the scale $\mu =M_f$ and with respect to $\mu =M_f =3$ GeV.
For both a) and b) the lines specified by + refer to the corresponding
quantities for $\sqrt{S}=100$ GeV.
\end{list}

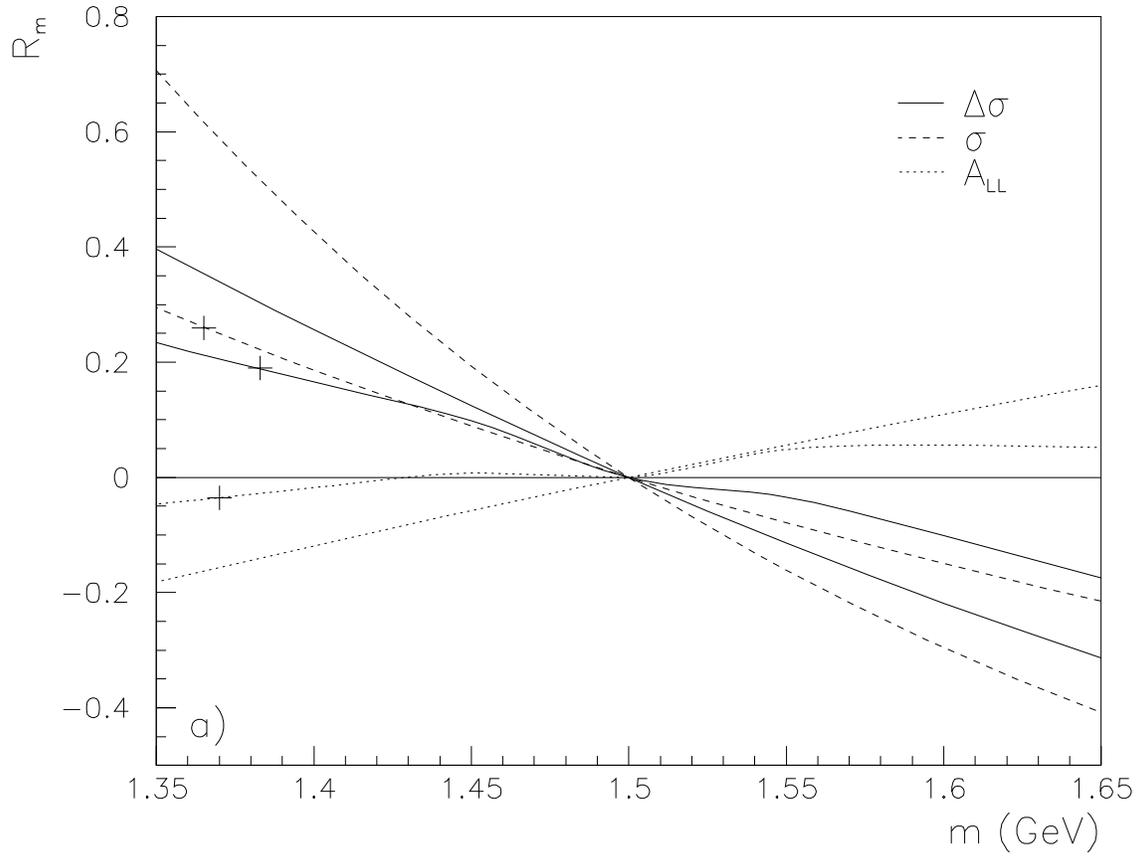
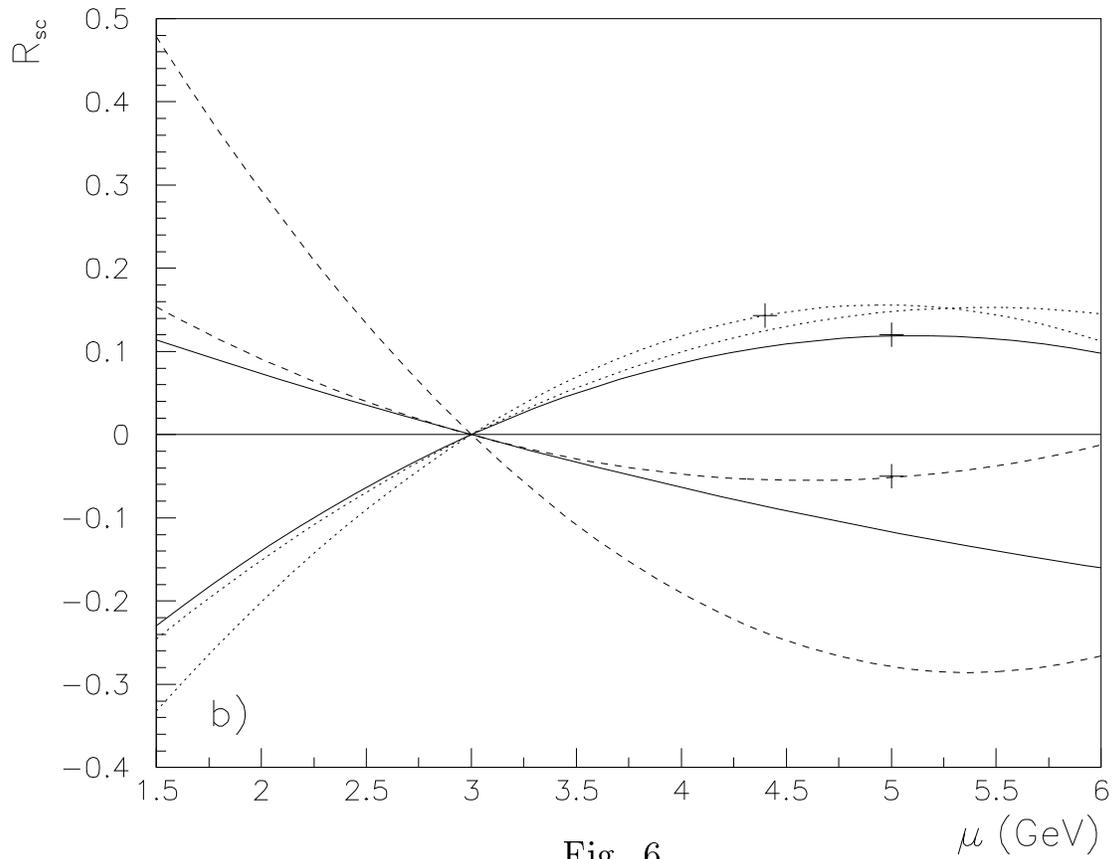
\end{document}